\documentclass[fleqn,usenatbib]{mnras}

\usepackage{newtxtext,newtxmath}

\usepackage[T1]{fontenc}

\DeclareRobustCommand{\VAN}[3]{#2}
\let\VANthebibliography\thebibliography
\def\thebibliography{\DeclareRobustCommand{\VAN}[3]{##3}\VANthebibliography}

\usepackage{graphicx}	
\usepackage{amsmath}	
\usepackage{xcolor}
\usepackage{ulem}

\title[Close galaxy pairs identification and analysis]{The PAU Survey: Close galaxy pairs identification and analysis}

\author[E. J. Gonzalez et al.]{ 
E. J. Gonzalez,$^{1,2,3}$\thanks{also at Port d'Informació Científica (PIC), Campus UAB, C. Albareda s/n, 08193 Barcelona (Barcelona), Spain.}\thanks{E-mail: ejgonzalez@unc.edu.ar} F. Rodriguez,$^{1,2}$\thanks{facundo.rodriguez@unc.edu.ar}
D. Navarro-Gironés,$^{4,5}$
E. Gaztañaga,$^{4,5,6}$
M. Siudek,$^{3,4}$
\newauthor{
D. Garc\'ia Lambas,$^{1,2}$ 
A. L. O’Mill,$^{1,2}$   
P. Renard,$^{7}$
L. Cabayol,$^{3}$\footnotemark[1]
J. Carretero,$^{3}$\footnotemark[1]
R. Casas,$^{4,5}$}
J. De Vicente,$^{8}$
\newauthor{
M. Eriksen,$^{3}$\footnotemark[1]
E. Fernandez,$^{3}$
J. Garcia-Bellido,$^{9}$
H. Hildebrandt,$^{10}$
R. Miquel,$^{3,11}$
C. Padilla,$^{3}$
E. Sanchez,$^{8}$}
\newauthor{
I. Sevilla-Noarbe,$^{8}$
P. Tallada-Crespí ,$^{8}$\footnotemark[1]
A. Wittje$^{10}$}
\\
$^{1}$Instituto de Astronomía Teórica y Experimental (IATE-CONICET), Laprida 854, X5000BGR, C\'ordoba, Argentina.\\
$^{2}$ Observatorio Astron\'omico de C\'ordoba, Universidad Nacional de C\'ordoba (OAC-UNC), Laprida 854, X5000BGR, C\'ordoba, Argentina.\\
$^{3}$Institut de Física d'Altes Energies (IFAE), The Barcelona Institute of Science and Technology, Campus UAB, 08193 Barcelona, Spain.\\
$^{4}$Institute of Space Sciences (ICE, CSIC), 08193 Barcelona, Spain.\\
$^{5}$Institut d'Estudis Espacials de Catalunya (IEEC), 08034 Barcelona, Spain.\\
$^{6}$ Institute of Cosmology \& Gravitation, University of Portsmouth, Dennis Sciama Building, Burnaby Road, Portsmouth PO1 3FX, UK\\
$^{7}$Department of Astronomy, Tsinghua University, Beijing 100084, China.\\
$^{8}$Centro de Investigaciones Energeticas, Medioambientales y Tecnologicas (CIEMAT), Avenida Complutense 40, E-28040 Madrid, Spain.\\
$^{9}$Instituto de Fisica Teorica UAM/CSIC, Universidad Autonoma de Madrid, Cantoblanco 28049 Madrid, Spain \\
$^{10}$Ruhr University Bochum, Faculty of Physics and Astronomy, Astronomical Institute (AIRUB), German Centre for Cosmological Lensing, 44780 Bochum, Germany\\
$^{11}$Institució Catalana de Recerca i Estudis Avançats (ICREA), 08010 Barcelona, Spain}

\pubyear{2022}

\begin{document}
\label{firstpage}
\pagerange{\pageref{firstpage}--\pageref{lastpage}}
\maketitle

\begin{abstract}
Galaxy pairs constitute the initial building blocks of galaxy evolution, which is driven through merger events and interactions. Thus, the analysis of these systems can be valuable in understanding galaxy evolution and studying structure formation.  In this work, we present a new publicly available catalogue of close galaxy pairs identified using photometric redshifts provided by the Physics of the Accelerating Universe Survey (PAUS). To efficiently detect them we take advantage of the high-precision photo$-z$ ($\sigma_{68}$ < 0.02) and apply an identification algorithm previously tested using simulated data. This algorithm considers the projected distance between the galaxies ($r_p < 50$\,kpc), the projected velocity difference ($\Delta V < 3500$\,km/s) and an isolation criterion to obtain the pair sample. We applied this technique to the total sample of galaxies provided by PAUS and to a subset with high-quality redshift estimates. Finally, the most relevant result we achieved was determining the mean mass for several subsets of galaxy pairs selected according to their total luminosity, colour and redshift, using galaxy-galaxy lensing estimates. For pairs selected from the total sample of PAUS with a mean $r-$band luminosity $10^{10.6} h^{-2} L_\odot$, we obtain a mean mass of $M_{200} = 10^{12.2} h^{-1} M_\odot$, compatible with the mass-luminosity ratio derived for elliptical galaxies. We also study the mass-to-light ratio $M/L$ as a function of the luminosity $L$ and find a lower $M/L$ (or steeper slope with $L$) for pairs than the one extrapolated from the measurements in groups and galaxy clusters.
\end{abstract}




\section{Introduction}

The current cosmological paradigm assumes that galaxies form by baryon condensation within the potential wells defined by the collisionless collapse of dark matter halos  \citep{WhiteRees1978}. These halos are expected to evolve according to the hierarchical formation scenario, in which smaller halos merge to form larger and larger ones in a ‘bottom-up’ fashion. In this context, galaxy pairs can provide valuable information about the early formation stages of more numerous systems. Particularly, close galaxy pairs (e.g., projected distances $r_\mathrm{p}$ $\leq 50\, h^{-1}$ kpc) are ideal places to study the interactions between galaxies that play a key role in galaxy evolution and their physical properties \citep{Toomre1972, Alonso2004,Hernandez2005,Alonso2007, Woods07, Ellison2010, Mesa2014,Patton2016}. 

Since these systems constitute the primordial bricks in galaxy evolution through merger events and interactions, galaxy pairs can also be studied to test different cosmological paradigms. It is important to take into account that galaxies do not evolve as isolated systems but as a part of the complex network that constitutes the cosmic web \citep{Bond1996}. Galaxy interactions can modify the mass distribution and morphology of galaxies and trigger star formation activity, varying in turn the observable galaxy properties \citep{Das2021,Garduno2021}. Therefore, the study of galaxy pairs can be important to better understand the colour distributions of observed galaxies, which is relevant in the calibration of cosmological tests that make use of subsets of galaxies targeted according to their colour, such as the Dark Energy Spectroscopic Instrument \citep[DESI; ][]{DESI,DESI2022}. Moreover galaxy pairs can be useful to test the $\Lambda$CDM paradigm, since obtaining the intervelocity distribution \citep{Pawlowski2022}, analysing the magnitude gap and its evolution \citep{Ostriker2019} or the study of the galaxy pair fraction can be used to constrain this cosmological scenario \citep{Robotham2014, huvko2022}.  

In view of the relevance of the galaxy pairs in the study of galaxy formation and cosmology, having a reliable sample of these systems is important as a starting point to analyse them in depth.
Galaxy pair catalogues are generally built considering a limiting velocity difference, $\Delta V$, computed according to spectroscopic redshifts, and a limiting projected distance between the member galaxies, $r_\mathrm{p}$ \citep{Lambas2003, Lambas2012, Robotham2014, Ferreras2017, Nottale2018}. However, this identification methodology requires spectroscopic information. One of the main challenges in cosmology today is obtaining large samples with accurate and precise redshifts. At present, high-resolution spectroscopy is inadequate to deal with the large volumes explored by modern wide-field galaxy surveys, given that this observational method is expensive in telescope time. 

Photometric redshift (photo$-z$) estimates thus emerge as an alternative to meaningfully expand the samples used to study the evolution of large-scale structure \citep[an updated description of photo$-z$ methods can be found in][]{Salvato2019}. In particular, the Physics of the Accelerating Universe Survey \citep[PAUS;][]{Benitez2009, Marti2014, Tonello2019, Padilla2019, Serrano2022} is a project designed for acquiring high-quality photo$-z$ measurements by combining the data of 40 narrow-band photometric filters with existing broad-band photometry. The low-resolution spectra from PAUS yield up to one order-of-magnitude improvements in the precision of photo$-z$, as compared with broad-band-only estimates \citep{Hildebrandt2012,Marti2014, Hoyle2018,Cabayol2019,   Eriksen2019, Alarcon2020, Cabayol2020, Cabayol2022, Soo2021, Alarcon2021}. The data products of this survey have been used in the analysis of many astrophysical process, such as the measurement of galaxy properties \citep{Tortorelli2021,Renard2022}, and cosmological studies such as the study of intrinsic alignments, clustering \citep{Johnston2021} and cosmic shear \citep{vandenBusch2022}. 

 Identifying galaxy pairs using only photometric information is difficult given the uncertainty of redshift estimates that leads to biased catalogues because of projection effects. However, \cite{Rodriguez2020} presented an algorithm to find close galaxy pairs using a high-precision photometric redshift catalogue ($\sigma_{68}$ < 0.02), which was assessed using MICE-GC\footnote{More details of this simulation can be found at \url{http://maia.ice.cat/mice/}.} simulation \citep{Fosalba2015, Carretero2015, Fosalba2015b, Crocce2015}. The results demonstrated that this procedure allows to identify these systems with a purity and a completeness of the sample higher than 0.8, successfully reproducing the distribution of total luminosity and mass of truly bound galaxy pairs residing in the same dark matter halo.


In addition to the galaxy systems identification, determining the masses of the dark matter halos in which they reside provides us with relevant information for understanding the connection between the baryonic and the dark matter content in these systems and the evolution of this relationship. Since galaxy pairs are an intermediate stage in the passage from halos containing one galaxy to those containing many-member systems, mass estimates of the host halo can contribute to the understanding of the joint evolution of galaxies and groups. This information is important, for example, in the Halo Occupation Distribution (HOD) context to link the average number of galaxies residing in a halo to its mass \citep[e.g.][]{Cooray2002, Zheng2005, Yang2007, Rodriguez2015, Rodriguez2020b}. 

One technique that has been demonstrated to be very efficient in obtaining the total mass of galaxy systems is weak gravitational lensing \citep[e.g.][]{Wegner2011,Dietrich2012,Jauzac2012,Umetsu2014, Jullo2014, Gonzalez2016, Gonzalez2018, Rodriguez2020,Gonzalez2021}. In particular weak-lensing stacking techniques have proved to be a powerful strategy that allows measuring the mean mass of the galaxy systems that are combined in the procedure \citep[e.g.][]{Leauthaud10, Melchior13,Rykoff08,Foex14,Chalela2017,Chalela2018, Pereira2018, Gonzalez2021}. These techniques boost the lensing signal by artificially increasing the density of lensed galaxies from which the estimators are obtained. Thus, they allow obtaining the mean mass of low-richness galaxy systems which are likely to reside in low-mass halos ($< 10^{13.5} M_\odot$) and from which the individual lensing signal is undetected. In particular, these techniques have been successfully applied in recovering the mean mass of galaxy pairs, finding general agreement with HOD predictions and other works that link mass to luminosity \citep{Gonzalez2019}. Moreover, in \cite{Rodriguez2020} it was shown that the total mass-luminosity relation of the identified galaxy pairs can be properly recovered using this technique in observational data.  

In this work we identify close pairs of galaxies using the photo$-z$ provided by PAUS.
We apply the identification algorithm assessed in \cite{Rodriguez2020} to obtain the pair catalogue and then we determine the mean masses for several subsets of the identified systems using weak-lensing stacking techniques. Finally, we also describe some characteristics of the close pairs of galaxies and the dependence of their mass on some of their properties, in particular we inspect the mass-luminosity ratio obtained for the analysed pair subsets. Our work is organised as follows: we describe PAUS photo$-z$ catalogue in Sec. \ref{sec:pau_data}; the identification procedure and the general properties of the pair catalogue are presented in Sec. \ref{sec:pairs}; we describe the lensing analysis and how the masses are determined for different galaxy pair subsets in Sec. \ref{sec:lens}; resultant masses are compared with the mean pair luminosity in Sec. \ref{sec:results} and finally we discuss and summarise our results in Sec. \ref{sec:conclusion}. We make publicly available our pair galaxy catalogue through the  the CosmoHub platform\footnote{\href{https://cosmohub.pic.es/home}{https://cosmohub.pic.es/home}} \citep{Carretero2017,TALLADA2020100391}. The description of the released data is described in appendix \ref{app:columns}.

\section{PAU Survey data }
\label{sec:pau_data}

PAUS is a narrow-band photometric wide-field galaxy survey conducted at the William Herschel Telescope in the Observatorio del Roque de Los Muchachos in La Palma (Canary Islands, Spain). The survey is carried out using the PAU camera \citep[PAUCam][]{Castilla2012, Castander2012, Padilla2016, Padilla2019}, an 18 CCDs camera that covers about 1 square degree of the sky, and reaches a limiting $i-$band magnitude of $\sim23$. The novelty of PAUCam resides in its set of 40 narrow-band filters spanning the range 4500-8500 Å with a width of 130 Å and equally spaced at 100 Å  \citep{Eriksen2019}.  

The target fields covered by PAUS are the COSMOS field \citep{Scoville2007}, the CFHTLS\footnote{CFHTLS catalog query page: \url{https://www.cfht.hawaii.edu/Science/CFHTLS/}} \citep{Cuillandre2012,Hudelot2012} ‘W1’, ‘W3’ and ‘W4’ fields and GAMA \citep{Driver2022} ‘G09’ field that overlaps the KiDS survey \citep{Kuijken2019}.  CFHTLenS \citep[][further description of this catalogue is provided in \ref{subsec:sources}]{Hildebrandt2012,Heymans2012} and KiDS catalogues are references from which PAUS forced photometry is performed \citep{Serrano2022}. Since they are deeper than PAUS, it ensures the completeness in the identification. The broad band magnitudes provided by these catalogues are also used in the photo$-z$ computation. In particular in this work, we use the data from the ‘W1’and ‘W3’ regions. We do not take into account COSMOS and W4 since they are too small (1.8 and 0.2 deg$^{2}$, respectively). In the case of G09, we discard this field from the analysis for consistency in the broad-band photometry. Moreover, the KiDS lensing catalogue is shallower than CFHTLS and the inclusion of the pairs identified in this field does not significantly improve the lensing analysis.

The photometric redshifts for the objects identified in each of these fields are computed via the code \texttt{BCNz2} \citep[][ Navarro-Gironés et al. in preparation]{Eriksen2019}, which is a template-based photometric redshift code. \texttt{BCNz2} is specifically designed to deal with the 40 narrow-bands filters of PAUS and the CFHTLS broad bands.  The final products of \texttt{BCNz2} are the photometric redshifts of the catalogue, the redshift probability distribution $p(z)$ and some photometric redshift quality parameters. For a more detailed description of the methodology applied to compute the photometric redshifts, refer to \cite{Eriksen2019}. With this procedure, PAUS achieves photometric redshifts with a precision of $\sigma (z)_\mathrm{RMS}$ ${\simeq }$ 0.0035(1 +$z$) for the 50\% of galaxies in the COSMOS field based on a photometric quality cut with $i$ < 22.5 \citep[][]{Marti2014, Eriksen2019}.

The performance of the photometric redshift can be quantified through the quality indicator, Qz, introduced by \citet{Brammer2008}:
\begin{equation} \label{eq:Qz}
    \textrm{Qz} \equiv \frac{\chi^{2}}{N_\mathrm{f}-3}\left (\frac{z_{\textrm{quant}}^{99} - z_{\textrm{quant}}^{1}}{\textrm{ODDS}} \right ),
\end{equation}
where $\chi^{2}$ is the chi-square of the fit to the SED templates, $N_\mathrm{f}$ is the number of used filters (narrow bands and broad bands) and $z_{\textrm{quant}}^{1}$ and $z_{\textrm{quant}}^{99}$ are the 1 and 99 percentiles of the posterior distributions $p(z)$, respectively. ODDS is another quality parameter which accounts for the $p(z)$ around the photometric redshift peak, $z_\mathrm{b}$, and it is defined as:
\begin{equation} \label{eq:ODDS}
    \textrm{ODDS} \equiv \int_{z_\mathrm{b}-\Delta z}^{z_\mathrm{b}+\Delta z}\mathrm{d}z \; p(z),
\end{equation}
where $\Delta z$ is an interval around the peak and it is set to $0.01$ to match the narrow PDFs in PAUS.  We choose Qz to  characterise the redshift quality because it is an hybrid quantity of other quality indicators, then it is useful to characterise the photo$-z$ performance that can be affected by many problems \citep[for further details see ][]{Brammer2008}

\begin{figure}
    \includegraphics[scale=0.41]{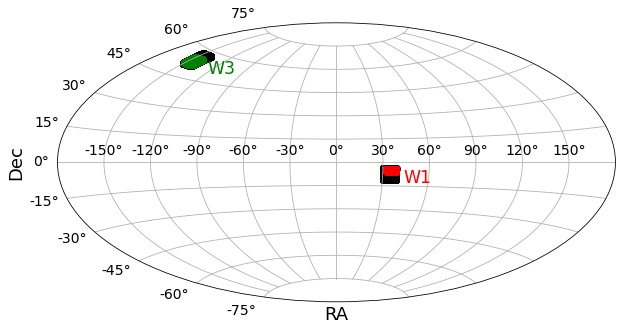}
    \caption{Spatial distribution of the 'W1' and 'W3' fields used for the analysis. In black, the positions of the sources used for the lensing analysis from the CFHTLenS catalogue (see \ref{subsec:sources}).}
    \label{fig:Area}
\end{figure}

Galaxy pairs are identified after removing objects considered as stars and bad quality objects according to the masks. We consider for the identification only the galaxies within a redshift range of $0.2 < z < 0.6$ following \citet{Rodriguez2020}, which is called the \textit{Total sample}. Figure \ref{fig:Area} shows the sky distribution of the galaxies from this sample in the wide fields used for the analysis. We also show the overlap with the sources selected for the lensing analysis (see \ref{subsec:sources}). Besides the \textit{Total sample}, we also consider a galaxy subset with a higher quality in the redshift estimates, named the \textit{Gold sample}. This sample is obtained by applying a cut in Qz, selecting the 25\% of the objects with better photometric redshifts. We also show in Table \ref{tab:sigma68_zb} the $\sigma_{68}$, the number of galaxies used for the analysis and the mean redshift for the samples analysed. $\sigma_{68}$ is defined as half of the difference between the two central quartiles of the $|z_{ \rm b}-z_{ \rm s}|/(1+z_{ \rm s})$ distribution, where $z_{ \rm s }$ are the spectroscopic redshifts used to validate the photo$-z$ performance, which were extracted from several surveys such as DEEP2 and VIPERS (for more details see Navarro-Gironés et al., in preparation). Obtained $\sigma_{68}$ are in agreement with the results presented in \citet{Eriksen2019} and are lower for the \textit{Gold sample} than for the \textit{Total sample}, as expected.

\begin{figure*}
    \includegraphics[scale=0.6]{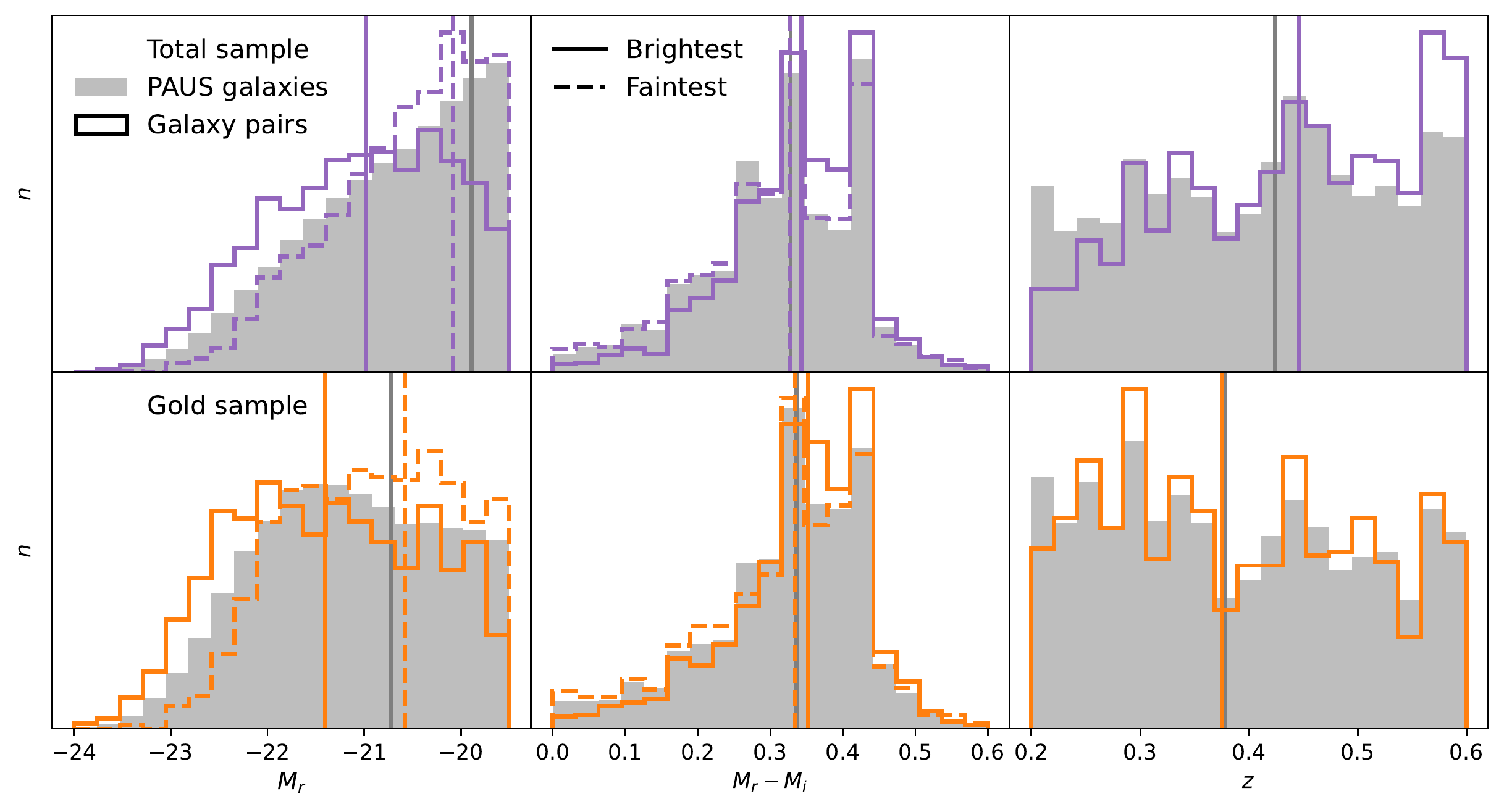}
    \caption{Normalised $r-$band absolute magnitude ($M_r$, left panels), colour ($M_r-M_i$, middle panels) and photo$-z$ ($z$, right panels) distributions for the galaxies in PAUS \textit{Total sample} (upper panels) and \textit{Gold sample} in solid grey bars. Unfilled distributions of magnitude and colours in solid and dashed lines correspond to the brightest and faintest galaxy members of the identified pairs, respectively. Unfilled distributions of redshift correspond to the redshift assigned to the identified pairs. Absolute magnitudes for PAUS galaxies and the pair members are computed considering the photo$-z$ provided by PAUS and the assigned pair redshift, respectively. Vertical lines indicate the median values, following the same colour and style code for each distribution.}
    \label{fig:pair-props}
\end{figure*}

For the pair identification we use CFHTLenS $r-$band photometry to perform the magnitude cuts and PAUS photo$-z$ estimates to compute $\Delta V$. The absolute magnitudes provided by the CFHTLenS have $k-$corrections computed using the CFHTLenS provided redshift applying the LePhare code \citep{Arnouts2002} and the technique presented by \citet{Ilbert2010}. Although this approach is not accurate since $k-$corrections are redshift-dependent, we do not expect this election to have a significant effect in the analysis, given that the introduced uncertainties due to the low redshift precision are expected to be large in the computation of these corrections. Nevertheless, to check that the CFHTLS magnitudes do not bias our results we computed absolute magnitude \texttt{CIGALE SED} fitting code \citep{Boquien2019} using PAUS photo$-z$ and narrow-band photometry complemented by CFHTLenS broad-band photometry \citep[][ Siudek et al. in preparation]{Renard2022,Tortorelli2021,Johnston2021}. We found a mean difference between CFHTLenS $k-$corrected magnitudes and PAUS absolute magnitudes of 0.04 which is less than typical error of the absolute magnitude estimations.


Fig. \ref{fig:pair-props} shows the distributions of the $r-$band absolute magnitude ($M_{r}$), the $r-i$ colour ($M_{r} - M_{i}$) and the photo$-z$ ($z$) for the \textit{Total sample} and the \textit{Gold sample}, together with the distributions for the identified pairs that will be better discussed in \ref{subsec:pairprops}. PAUS galaxies included in the \textit{Gold sample} are more luminous objects and located at lower redshifts than those in the \textit{Total sample}. This is expected since the determination of the photo$-z$ using \texttt{BCNz2} is more efficient for brighter and low redshift objects than for fainter and deeper ones.

\begin{table}
\caption{$\sigma _{68}$ of the quantity $|z_\mathrm{b} - z_\mathrm{s}| / (1 + z_\mathrm{b})$, number of galaxies and mean redshift of the samples in each field. }
\begin{center}
\begin{tabular}{|l c c c c|}
\hline
\hline
        & Field  &  $\sigma_{68}$ & $N_\text{gal}$ & $\langle z \rangle$ \\ \hline
\text{Total sample} & \text{W1} & 0.019 & 94687 & 0.404 \\
                    & \text{W3} & 0.020 & 221316 & 0.414 \\ \hline
\text{Gold sample}  & \text{W1} & 0.0028 & 33222 &  0.380\\ 
                    & \text{W3} & 0.0029 & 74694  & 0.392 \\
\hline
\end{tabular}
\end{center}
\label{tab:sigma68_zb}
\end{table}

\section{Close Galaxy pairs}
\label{sec:pairs}
\subsection{Galaxy pair identification}

In order to identify the galaxy pairs, we implement the procedure presented in \cite{Rodriguez2020}, which is based on a similar approach as those based on spectroscopic surveys but considering the uncertainties of the photometric redshift estimates. This algorithm was assessed using galaxy mocks taking into account the expected uncertainties in high-precision photometric redshift surveys like PAUS. The galaxy pairs identification is based on the traditional approach of linking galaxies that are close according to the projected distance, $r_\mathrm{p}$, and with a limiting velocity difference, $\Delta V$, computed according to PAUS photometric redshifts. This is performed by constraining the parameters $r_\mathrm{p}$ and $\Delta V$, respectively. Furthermore, the algorithm considers that the pair has at least one galaxy brighter than a certain magnitude threshold, impose a limiting magnitude difference between the members ($\Delta m$) and applies an isolation criterion. These combined criteria ensure the identification of galaxy pairs whose members have comparable masses discarding, in turn, those that are part of a larger system.

\begin{figure}
\begin{center}
    \includegraphics[scale=0.25]{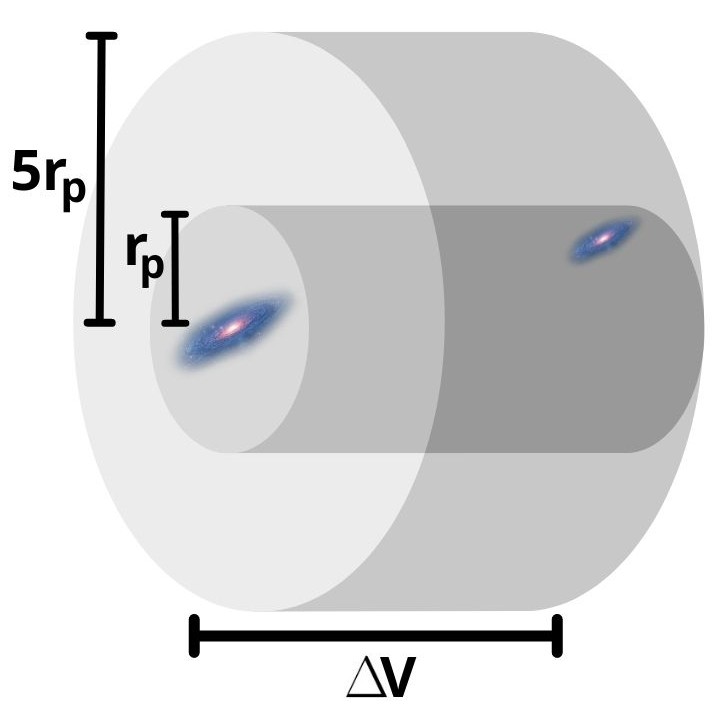}
    \caption{Galaxy pairs identification schema. For implementing the algorithm in this work, we set the values of $r_\mathrm{p}$ = 50 kpc and $\Delta v=3500$ km/s. In addition, the pairs must have at least one galaxy brighter than -19.5 in absolute magnitude of SDSS $r-$band and a magnitude difference between their members $\Delta m < 2$. The schema is only illustrative and the proportions are not to scale.}
    \label{fig:Schema}
    \end{center}
\end{figure}

\begin{figure*}
    \centering
    \includegraphics[width=0.66\columnwidth]{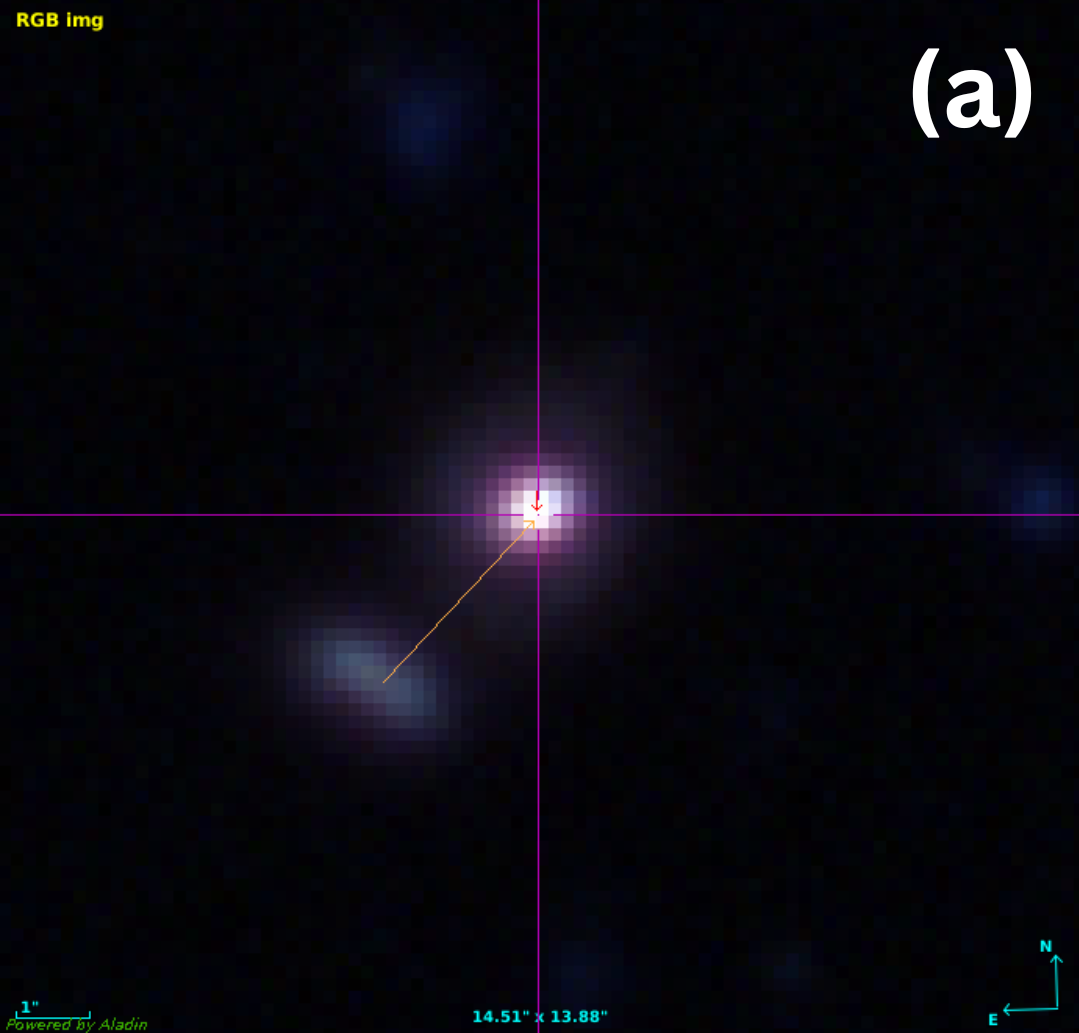}
    \includegraphics[width=0.66\columnwidth]{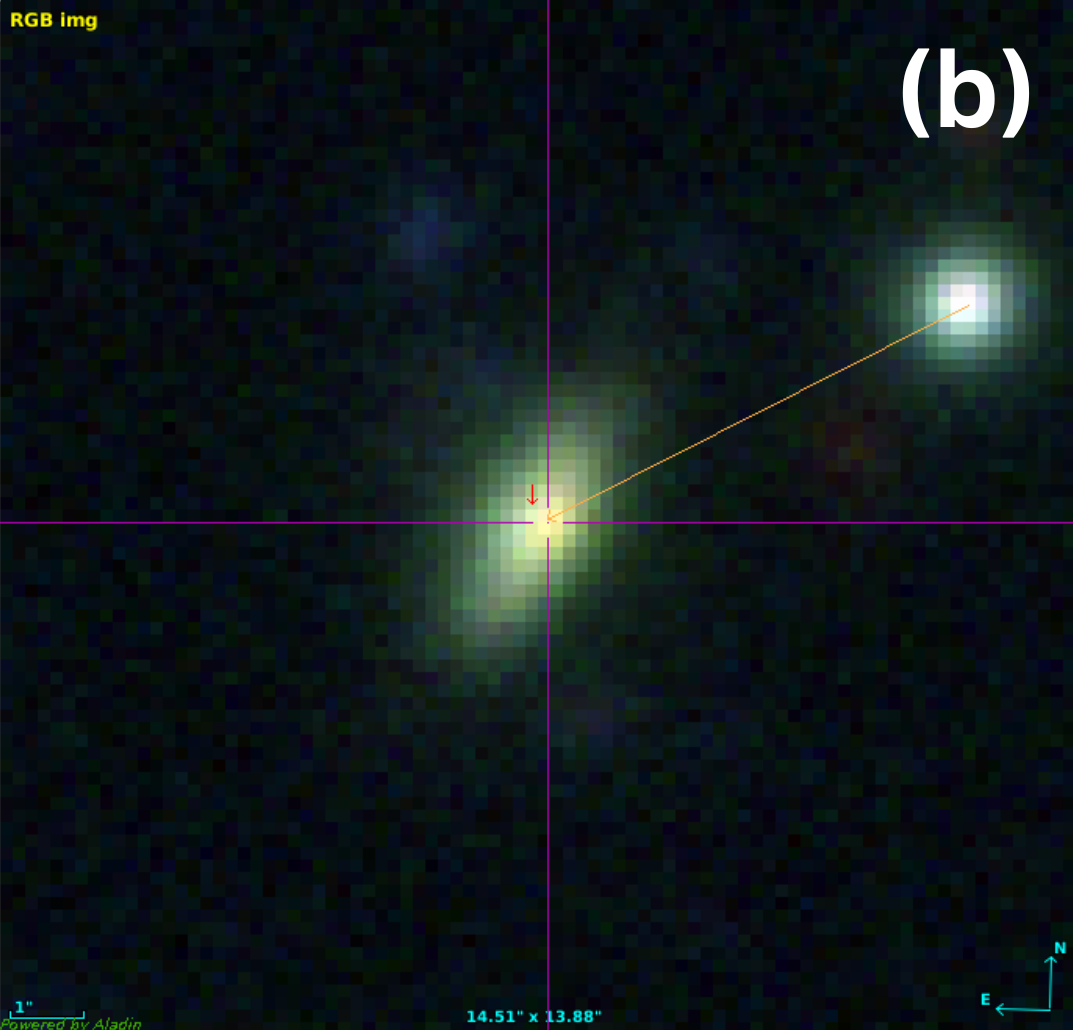}
    \includegraphics[width=0.66\columnwidth]{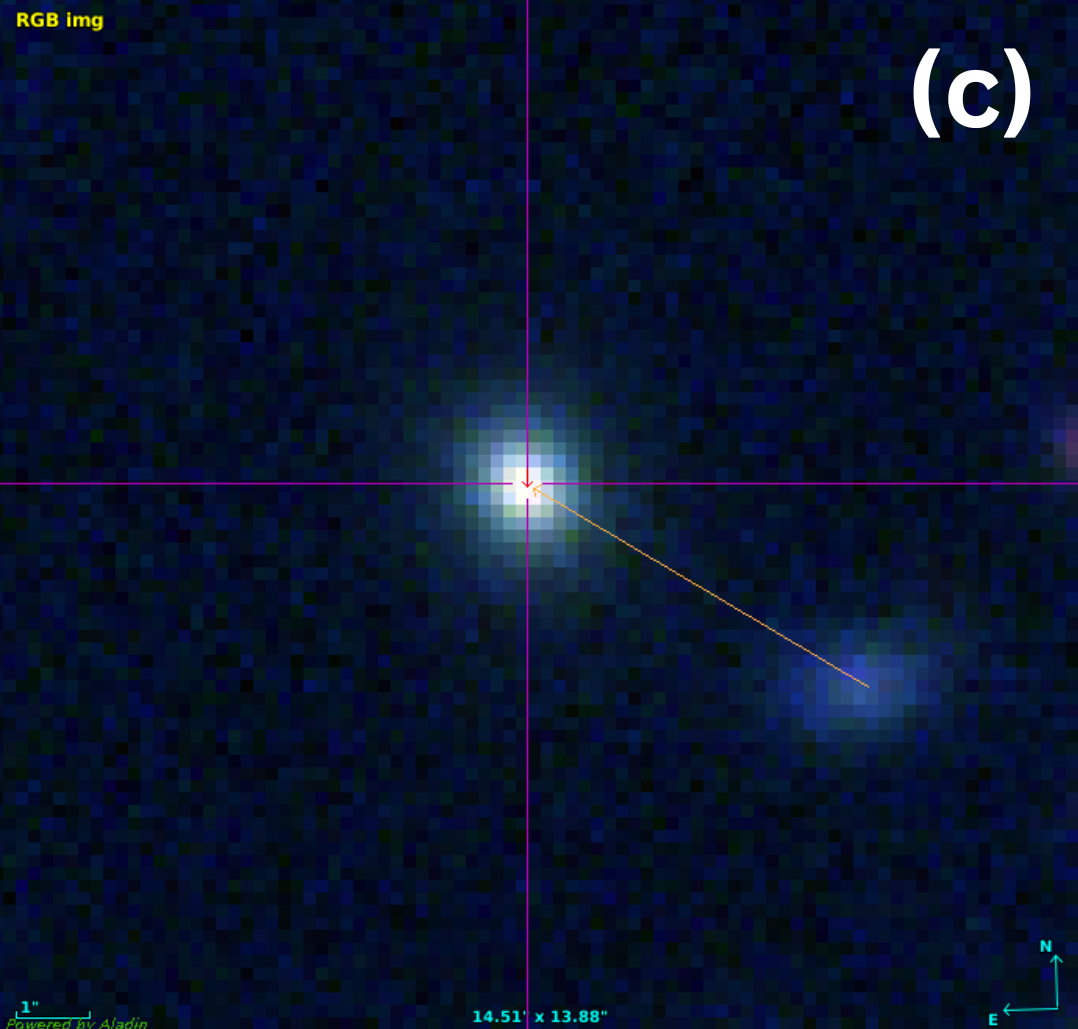}
    \includegraphics[width=0.66\columnwidth]{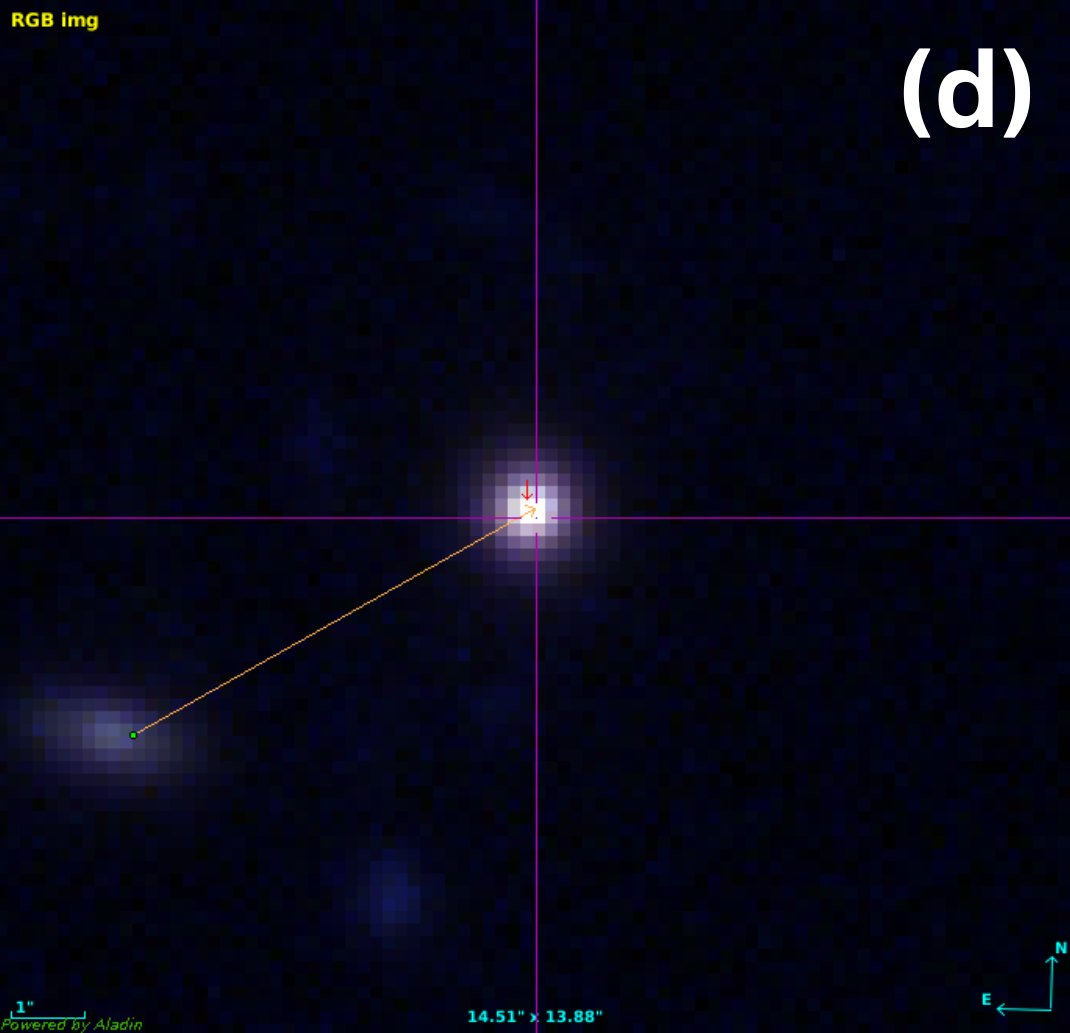}
    \includegraphics[width=0.66\columnwidth]{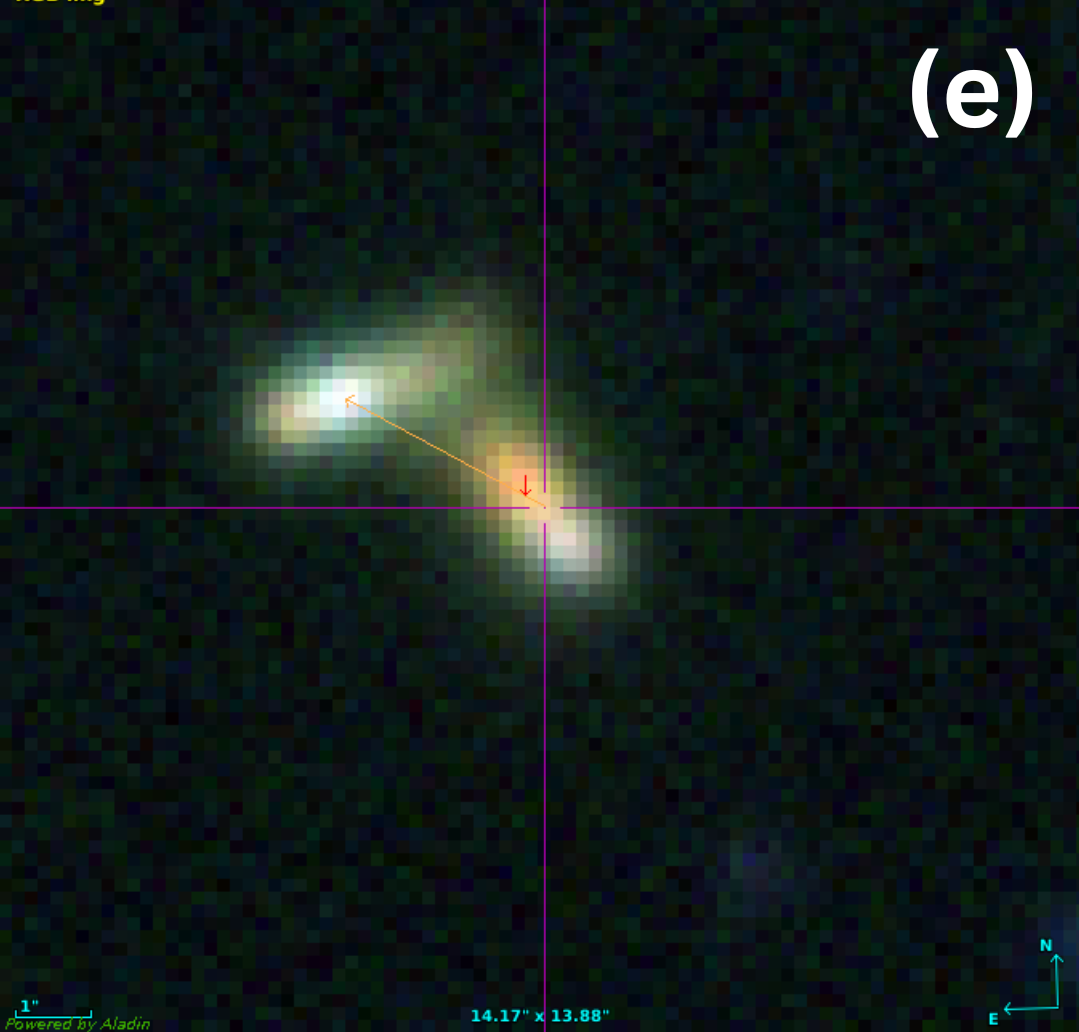}
    \includegraphics[width=0.66\columnwidth]{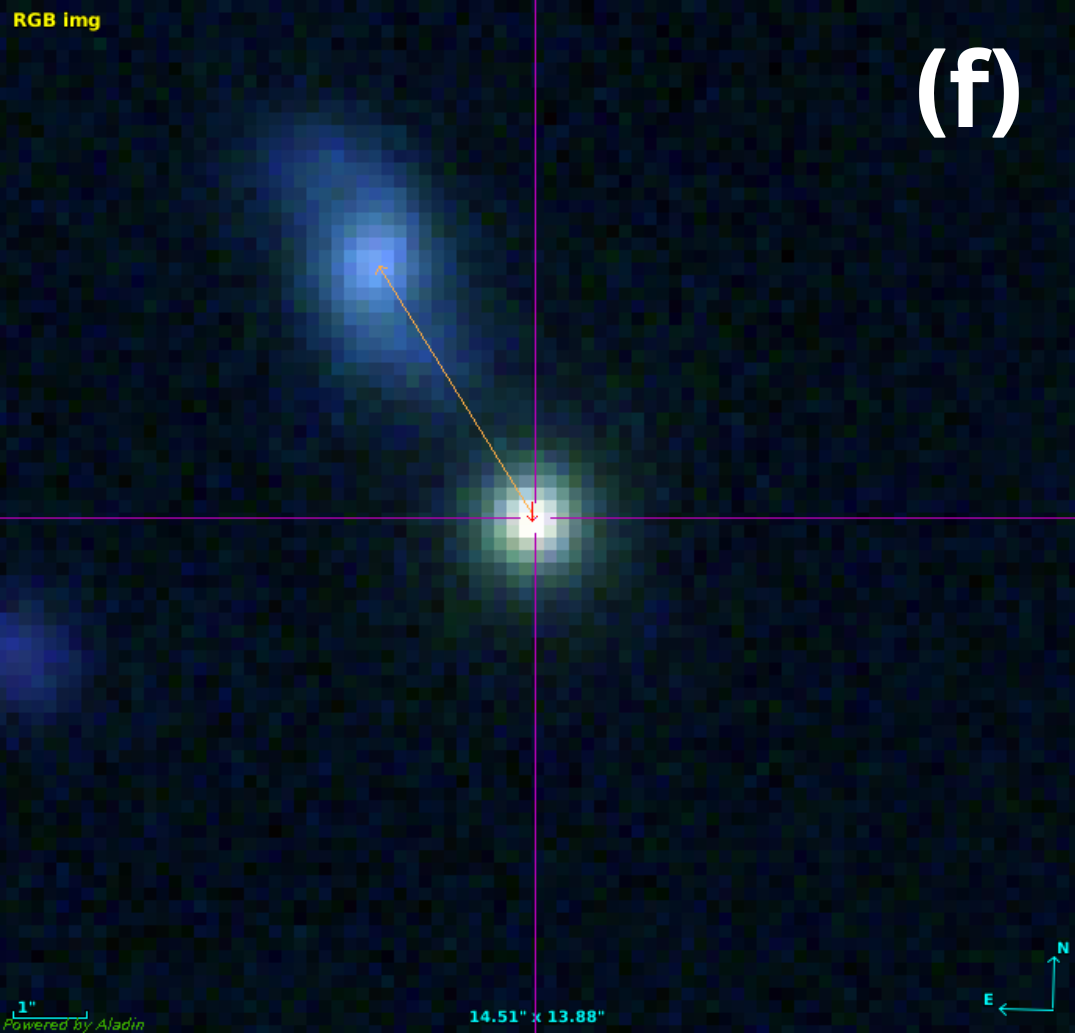}
    \caption{Examples of six identified galaxy pairs. Each panel is a 0.004-degree sideways RGB image resulting from combining $i-$, $g-$, and $r-$band data frames. The correspondent photometric redshift, $z_{pair}$, projected angular, $d$, and physical, $r_\mathrm{p}$, distances are: (a) $z_{pair} = 0.589$, $d = 3.044''$, $r_p=32.85 h^{-1}$kpc, (b) $z_{pair} = 0.395$, $d = 6.424''$, $r_p = 47.10 h^{-1}$kpc, (c) $z_{pair} = 0.513$, $d = 5.243''$, $r_p=49.36 h^{-1}$kpc, (d) $z_{pair} =0.381$, $d = 6.426''$, $r_p = 46.54 h^{-1}$kpc, (e) $z_{pair} = 0.566$, $d = 2.892''$, $r_p=27.02 h^{-1}$kpc; (f) $z_{pair} = 0.454$, $d=3.933''$, $r_p = 34.38  h^{-1}$kpc}
    \label{fig:pairimages}
\end{figure*}

\begin{table} 
 \caption{Number of galaxy pairs in each sub-sample selected according to their physical properties.}
    \centering
    \begin{tabular}{c c c }
    \hline
    \hline
Sub-sample   & Total sample & Gold sample   \\
\hline
all pairs & 2282 & 1114 \\
$M^\text{pair}_r < -22.5$ & 338 & 326\\
$M^\text{pair}_r \geq -22.5$ & 1944 & 788 \\
$z < 0.4$ & 852 & 588 \\
$z \geq 0.4$ & 1430 & 526\\
$L_2/L_1 < 0.5$ & 1251 & 542\\
$L_2/L_1 \geq 0.5$ & 1031 & 572\\
blue pairs & 1103 & 534\\
red pairs & 1179 & 580 \\
\hline
\end{tabular}
\begin{flushleft}
\end{flushleft}
\label{tab:sampdef}
\end{table}

\begin{figure*}
    \includegraphics[scale=0.6]{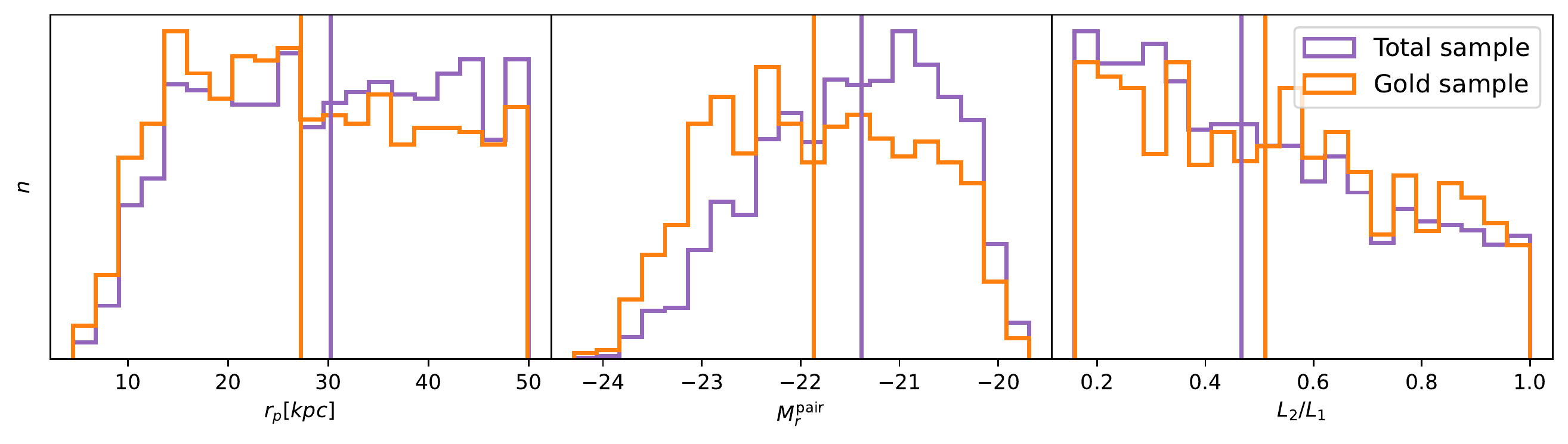}
    \caption{Normalised distributions of the projected distances, $r_\mathrm{p}$, absolute total $r-$band magnitudes, $M^\text{pair}_r$, and luminosity ratio, $L_2/L_1$, for the pairs in the \textit{Total} and \textit{Gold samples} of galaxies (purple and orange solid lines, respectively).}
    \label{fig:pair-lratio}
\end{figure*}

\begin{figure}
    \includegraphics[scale=0.6]{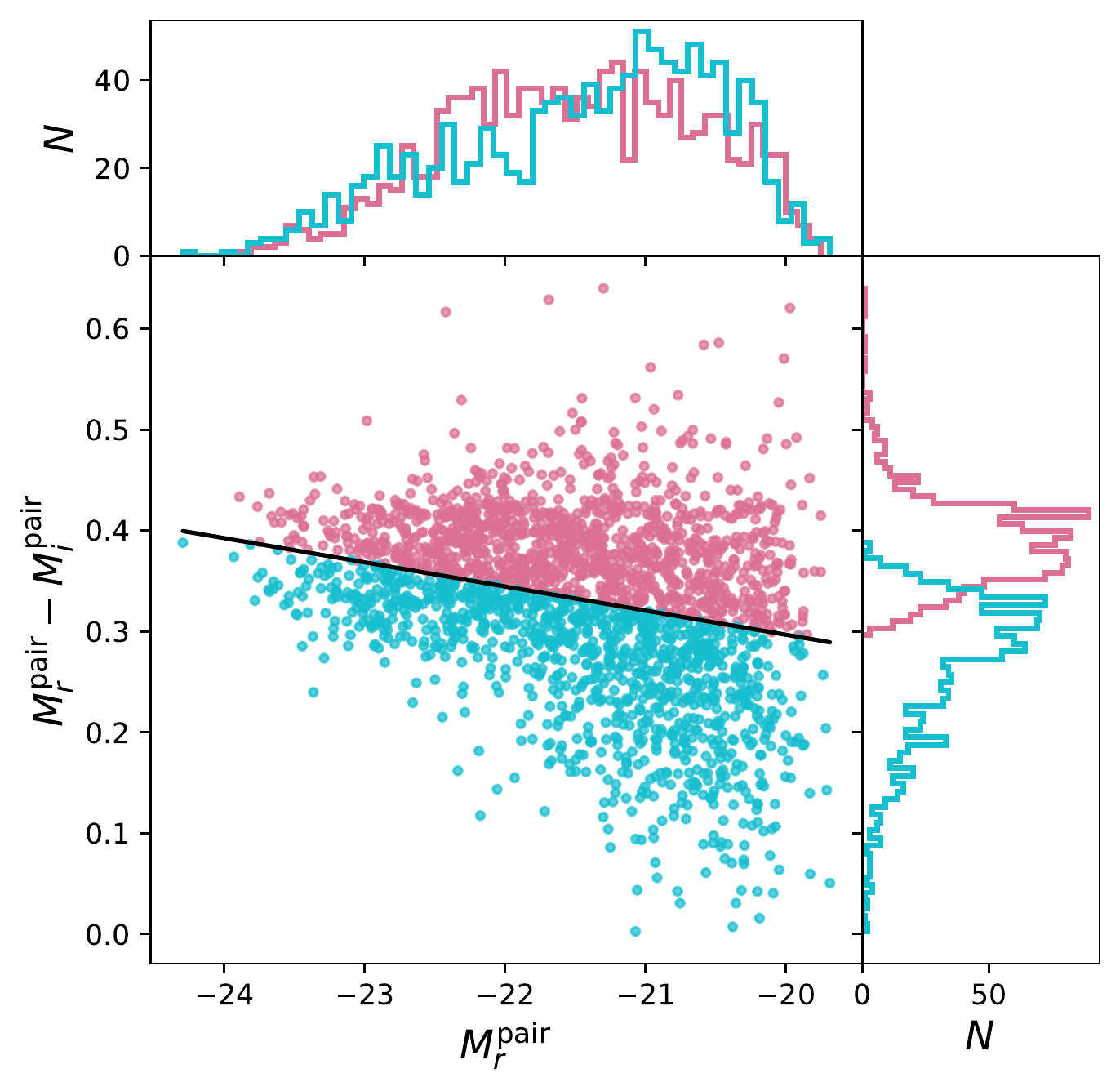}
    \caption{Colour-magnitude diagram for the pairs selected from the \textit{Total sample} of galaxies. The black solid line is a linear fit of the pair median colours in $M^\text{pair}_r$ bins. Pink and light-blue dots correspond to the red and blue pair sub-samples, respectively. Lateral and upper panels show the distributions of colour and magnitude of these sub-samples. }
    \label{fig:color}
\end{figure}

The procedure starts with the selection of possible pair centres, by considering only the galaxies brighter than a $r-$band absolute magnitude of $M_r - 5\log h = -19.5$. Then, we search for another galaxy fainter than the potential centre, within a projected radius $r_\mathrm{p}$ = 50 kpc and a given velocity difference, $\Delta V$, that depends on the photometric redshift error. Considering the uncertainties in the photo$-z$ samples used and following \cite{Rodriguez2020}, we set a criterion of $\Delta V \leq 3500$ km/s. Besides, the sample is restricted to those systems that also satisfy an apparent magnitude difference of $\Delta m < 2$ between their members. Finally, an isolation criterion is applied by requiring that no other galaxy lies within 5$\, r_p$. A schema of the identification is shown in Fig \ref{fig:Schema}. The assigned redshift to the pair system, $z^\text{pair}$, is computed as the average of the galaxy member photometric redshifts. Then, we recompute absolute magnitudes using PAUS photo$-z$ and then considering the assigned redshift to the pair after the identification to obtain the total luminosity of the system.    

After implementing the algorithm described above, we obtain 2282 pairs using the PAUs \textit{Total sample} of galaxies and 1114 pairs identified using the \textit{Gold sample}. To illustrate how the identified systems look like, we show images of six found galaxy pairs from the \textit{Total sample} in Fig. \ref{fig:pairimages}. Each panel is a 0.004 degree RGB image composed of $i-$, $g-$ and $r-$band data taken from the Hyper Suprime-Cam Subaru Strategic Program Public Data Release\footnote{The images were downloaded from \url{https://hsc-release.mtk.nao.ac.jp/}.} that overlaps with the `W1' field. Taking into account the identification process, they are all close pairs. These images help us to exemplify this and to be more confident in the identification procedure. It allows us to see that some systems, such as (e) and (f), show signs of interaction between the members.

\subsection{Galaxy pair properties}
\label{subsec:pairprops}
In Fig. \ref{fig:pair-props} we show the distributions of $r-$band magnitudes, $r-i$ colours and photometric redshifts of PAUS \textit{Total} and \textit{Gold sample} of galaxies (solid grey distributions), together with the distributions for the galaxy pair members identified in each sample (unfilled colour distributions). Similar properties such as the observed for the complete sample of galaxies are obtained for the members of the galaxy pairs identified.

In Fig. \ref{fig:pair-lratio} we show the total absolute magnitude in the $r-$band, $M^\text{pair}_r = -2.5 \log{(L_1 +  L_2)} $, where $L_2$ and $L_1$ correspond to the $r-$band luminosity of the faintest and the brightest galaxy of the pair system, and the luminosity ratio of the members of identified pairs, $L_2/L_1$. Pairs from \textit{Gold sample} are more luminous and include a lower fraction of pairs with lower $L_2/L_1$, i.e. higher $\Delta m$ values, than those identified using the \textit{Total sample} of galaxies. Our derived luminosity ratios correspond to a mean magnitude gap, $\langle \Delta m \rangle$, of $0.88$ and $0.83$ for the pairs in the \textit{Total} and the \textit{Gold sample}, respectively. This gap is lower than observed for galaxy systems in simulations \citep[$\langle \Delta m \rangle > 1.5$][]{Ostriker2019} given that we are imposing a limiting gap of $2.0$ through the identification. Moreover, we are identifying close galaxy pairs of similar magnitudes, thus we are mainly detecting the systems in the pre-merging phase of the main component with the secondary one, which is one of the main sources that is expected to increase the observed gap. 


Pairs are also classified as \textit{red} and \textit{blue} according to their position in the colour-magnitude diagram, as shown in Fig. \ref{fig:color}. We split the sample of pairs into ten percentiles of absolute magnitude and compute the median colour for each bin. Then, we fit a linear function to the obtained median values. \textit{Red} (\textit{blue}) pairs are those with higher (lower) colours than the values given by the fitted function. With this procedure, we obtain pair sub-samples classified according to the colour with similar total luminosity distributions. We highlight that this classification, in fact, selects the redder and bluer pairs of the obtained catalogue which may not be complete or do not have balanced colours. We take into account the $r-i$ colour instead of other more commonly used colour definitions, such as $g-r$ \citep[e.g.,][]{Patton2011,Rodriguez2015,Gonzalez2018} or $u-r$ \citep[e.g.,][]{Perez2005,Das2022} since our analysis includes higher redshift systems than those included in previous works. A further inspection of the pair properties related with the colour will be performed in a following work.

To analyse the dependence between the total mass of the systems and their properties, we split the samples into sub-sets as specified in Table \ref{tab:sampdef}. The cuts selected are also inspired by the previous study presented in \citet{Rodriguez2020}. We modify the threshold adopted for the total magnitude and luminosity ratio sub-sets since the distributions differ significantly from the predicted according to the simulated data. This can be the result of a lack of the mock catalogue in reproducing the properties of these particular systems. However, the predicted mass-luminosity ratio for the analysed pairs is in agreement with the expected according to the simulated data (see \ref{subsec:mlrelation}).

\section{Lensing mass estimates}
\label{sec:lens}
\subsection{Lensing analysis}

The gravitational lensing effect is generated by the presence of a gravitational potential, called lens, which bundles the light of the sources that are behind this potential. This effect introduces a distortion in the shapes of extended luminous sources, such as background galaxies, that can be related to the projected surface density distribution of the lens, such as a galaxy system. In particular, the effect introduced at larger distances from the galaxy system centre, which is related to the weak-lensing regime, traces the mass distribution at the outskirts and provides information on the total halo mass content that hosts the galaxy system. The introduced distortion in the shapes of background galaxies can be quantified by their measured ellipticity, $\epsilon$, and related to the complex \textit{shear} parameter, $\gamma$. 

However, the main drawback of weak-lensing studies is that the observed distortion is very small and is combined with the intrinsic galaxy shape, thus, it can only be accounted for using a statistical approach.  Essentially, the strategy consists of averaging the ellipticities of many sources at roughly the same distance from the galaxy system centre, considering a radial isotropic mass distribution. By assuming that the intrinsic ellipticities are randomly orientated, the averaged ellipticity parameter is related only with the \textit{shear}, $\langle e \rangle = {\gamma}$, which is in turn related to the contrast density distribution, $\Delta\Sigma$, through \citep{Bartelmann1995}:
\begin{equation} \label{eq:DeltaSigma}
    \gamma_{\rm{t}}(r) \times \Sigma_{\rm crit} = \bar{\Sigma}(<r) -  \bar \Sigma(r)  \equiv \Delta \Sigma(r).
\end{equation}
Here $\gamma_{\rm{t}}(r)$ is the tangential component of the \textit{shear} at a projected distance from the centre of the mass distribution, $r$, $\bar{\Sigma}(<r) $ and $\bar \Sigma(r)$ are the azimuthally averaged projected surface density distribution within a disk and within a ring of distance $r$, respectively. $\Sigma_{\rm crit}$ is the critical density which can be obtained according to the observer-source-lens distances as:
\begin{equation} \label{eq:sig_crit}
\Sigma_{\rm{crit}} = \dfrac{c^{2}}{4 \pi \text{G}} \dfrac{D_\text{OS}}{D_\text{OL} D_\text{LS}},
\end{equation}
where $D_\text{OL}$, $D_\text{OS}$, and $D_\text{LS}$ are  the angular diameter distances from the observer to the lens, from the observer to the source and from the lens to the source, respectively.
On the other hand, the averaged cross-component of the shear, $\gamma_{\times}$, defined as the component tilted at $\pi$/4 relative to the tangential component, should be zero and commonly used as a null test to check for the presence of systematics in the data.

One of the main sources of noise introduced in the analysis is known as `shape noise' and it is proportional to the inverse square root of the number of source galaxies, $N_S$. This introduced noise can be lowered by applying stacking techniques, which artificially increases $N_S$ by combining many lenses. This allows us to derive an average projected density distribution of the combined lenses with a higher signal-to-noise ratio. The details on this procedure and how it can be applied to derive the total masses of galaxy pairs are presented in \citet{Gonzalez2019} and \citet{Rodriguez2020}. Here we briefly describe the lens catalogue used for the analysis and the fitting procedure applied to derive the masses. 

\subsubsection{Source galaxy catalogue and contrast density profile}
\label{subsec:sources}

To perform the lensing analysis we use the public CFHTLenS weak lensing catalogues. These catalogues are based on observations provided by the multiband survey ($u^*,g',r',i',z'$), CFHT Legacy Survey. The galaxy catalogue includes the shape measurements obtained using the \textit{lens}fit algorithm \citep{Miller2007,Kitching2008} applied on the $i-$band, with a resultant weighted galaxy source density of $\sim 15.1$\,arcmin$^{-2}$. It also includes photometric redshifts estimated using the BPZ algorithm \citep{Benitez2000,Coe2006}.
 See \citet{Hildebrandt2012,Heymans2012,Miller2013,Erben2013} for further details regarding this shear catalogue. CFHTLS spans over 154\,deg$^2$ distributed in four separate patches: W1, W2, W3 and W4 ($63.8$, $22.6$, $44.2$ and $23.3$ deg$^2$, respectively). W1 and W3 fully overlaps W1 and W3 PAUS fields (see Fig. \ref{fig:Area}).

For our analysis, we have only included galaxies considering the following \textit{lens}fit parameters cuts: 
MASK $\leq$ 1, FITCLASS $= 0$ and $w > 0$.
Here MASK is a masking flag, FITCLASS is a flag parameter that is set to $0$ when the source is classified as a galaxy and
$w$ is a weight parameter that takes into account errors on the shape measurements and the intrinsic shape noise \citep[see details in ][]{Miller2013}. Our lensing study is performed following \citet{Miller2013}, by applying the additive calibration correction factors for the ellipticity components provided for each catalogue and a multiplicative shear calibration factor to the combined sample of galaxies.

For each galaxy pair located at a photometric redshift $z^\text{pair}$, we select the sources taking into account
Z\_BEST $ > z^\text{pair} + 0.3$ and ODDS\_BEST $> 0.5$, where
Z\_BEST is the photometric redshift estimated for each galaxy,
and ODDS\_BEST is a parameter that expresses the quality of 
Z\_BEST and takes values from 0 to 1. In this selection, we only include those galaxies with Z\_BEST up to $1.2$.

The stacked contrast density profile is obtained by combining the sample of analysed pairs as:
\begin{equation}
 \langle \Delta \Sigma \rangle (r) = \frac{\sum_{j=1}^{N_\mathrm{L}} \sum_{i=1}^{N_\mathrm{S,j}} \omega_\mathrm{LS,ij} \Sigma_{{\rm crit},ij} e_{{\rm t},ij}}{\sum_{j=1}^{N_\mathrm{L}} \sum_{i=1}^{N_\mathrm{S,j}} \omega_\mathrm{LS,ij}},
\end{equation}
where $\omega_\mathrm{LS,ij}$ is the inverse variance weight computed according to the weight, $\omega_{ij}$, given
by the $lens$fit algorithm for each background galaxy, $\omega_\mathrm{LS,ij}=\omega_{ij}/\Sigma^2_{{\rm crit},ij}$. 
$N_\mathrm{L}$ is the number of galaxy pairs considered for the stacking and 
$N_\mathrm{S,j}$ the number of background galaxies located at a distance 
$r \pm \delta r$ from the $j$th pair. 
$\Sigma_{{\rm crit},ij}$ is the critical density for the $i-$th background galaxy of the $j-$th pair. We obtain the profiles by binning the background galaxies in 15 non-overlapping log-spaced $r$ bins, from $100 h^{-1}$kpc up to $10 h^{-1}$Mpc. For each radial bin errors are computed by bootstrapping the lensing signal using 100 realisations.

\begin{figure}
    \includegraphics[scale=0.6]{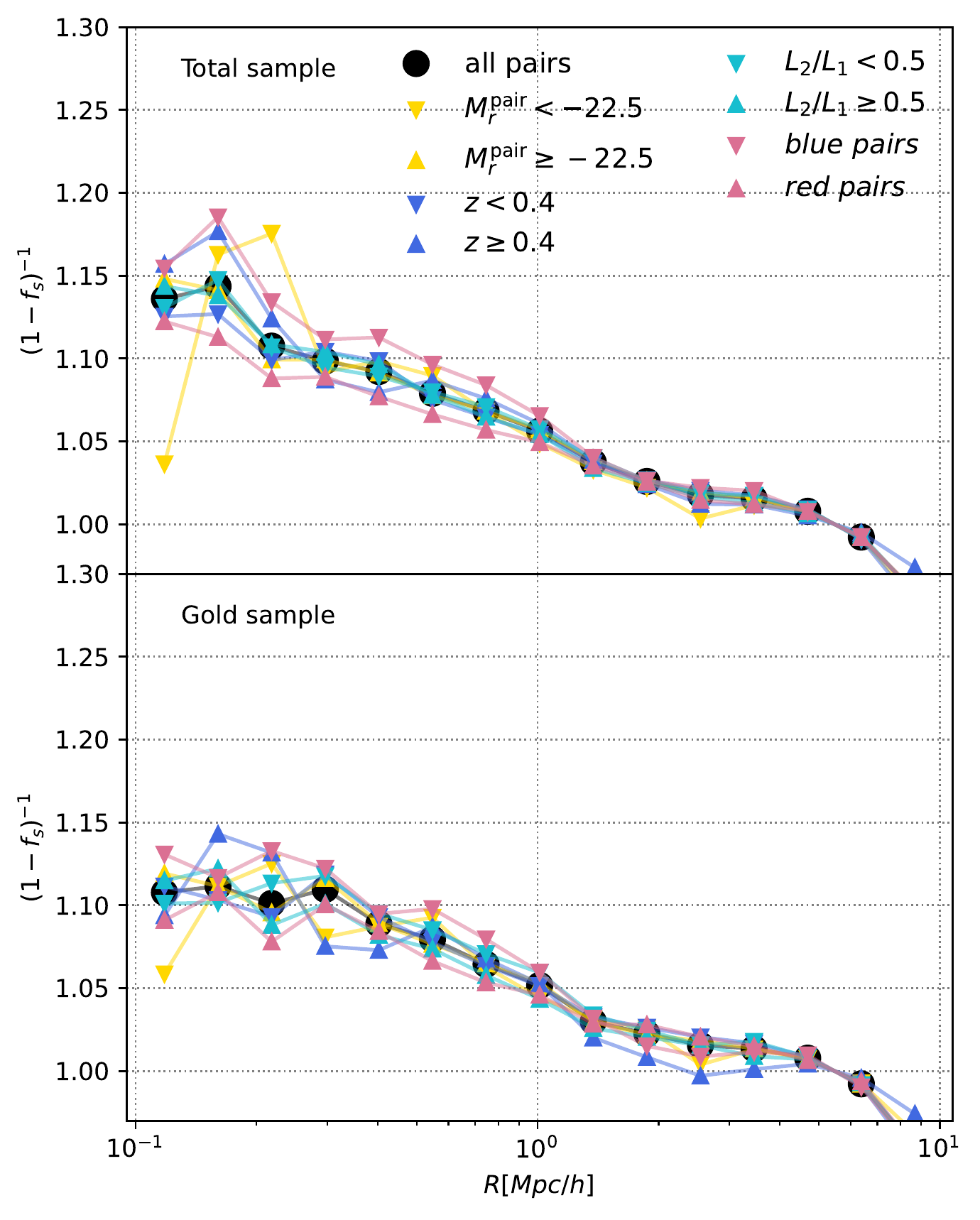}
    \caption{Boost factor derived according to the contamination fraction, $f_s$, computed as the excess in the radial density distribution of the background galaxies selected for the stacked pair sub-samples.}
    \label{fig:fs}
\end{figure}

Errors in the photometric redshifts can lead to the inclusion in the background galaxy sample of foreground or galaxy satellites that belong to the pair system. These galaxies are unlensed and result in an underestimated density contrast, which is called the dilution effect. In order to take this effect into account, the $\Delta \tilde{\Sigma}$ measurement can be boosted to recover the corrected signal by using the so-called \textit{boost-factor} \citep[e.g.,][]{Kneib2003, Applegate2014, Simet2017, Leauthaud2017, Melchior2017, Varga2019}: $1/(1-f_{s})$, where $f_{s}$ is the contamination fraction and it is expected to be higher in the inner radial bins where the contamination by low brightness satellites is more significant. We compute $f_{s}$ by using a similar approach as the one presented in \citet{Hoekstra2007}. Since a non-contaminated background galaxy sample will present a constant density for all the considered radial bins, by computing the excess in the density at each considered radial bin we obtain an estimated value of $f_{s}$. This excess is computed taking into account the background galaxy density obtained for the average of the 2 last radial bins, where the contamination of unlensed galaxies is expected to be negligible. By doing so, we obtain the $f_{s}(r)$ fraction, which is included in the analysis. In Fig. \ref{fig:fs} we show the radial boost factor computed for each stacked galaxy pair sub-sample. In general, there is an excess in the density of background galaxies of about $10\%$. This is indicating that there are other fainter satellites associated with the pair system identified. 

We also take into account a noise bias factor correction as suggested by \citet{Miller2013}, which considers the multiplicative shear calibration factor $m(\nu_\mathrm{SN},l)$ provided by \textit{lens}fit, where $\nu_\mathrm{SN}$ and $l$ are the signal-to-noise of the shape measurement and the shape of the galaxy, respectively. We compute:
\begin{equation}
1+K(z_\mathrm{L})= \frac{\sum_{j=1}^{N_\mathrm{L}} \sum_{i=1}^{N_{\mathrm{S},j}} \omega_{\mathrm{LS},ij} (1+m(\nu_{\mathrm{SN},ij},l_{ij}))}{\sum_{j=1}^{N_\mathrm{L}} \sum_{i=1}^{N_\mathrm{S,j}} \omega_{\mathrm{LS},ij}}
\end{equation}
and following \citet{Velander2014,Hudson2015,Shan2017,Leauthaud2017,Pereira2018} we compute the corrected profile multiplying them by a 
factor $(1+K(z_\mathrm{L}))^{-1}$. 

Taking the corrections into account we obtain the calibrated estimator as:
\begin{equation} \label{eq:estimators}
 \text{E} \left( \frac{f_s(r)  \langle \Delta \Sigma \rangle (r)}{1+K(z_\mathrm{L})} \right)  = \Delta \Sigma(r), 
\end{equation}

\subsubsection{Modelling and fitting procedure}

We model the derived contrast density distributions by considering a sum of two terms: 
\begin{equation} \label{eq:dsigma}
\Delta \Sigma = \Delta \Sigma_{1 \rm h} + \Delta \Sigma_{2 \rm h}.
\end{equation}
$\Delta \Sigma_{1 \rm h}$ is related with the main halo component, while $\Delta \Sigma_{2\rm h}$ is called the 2-halo term component and is related with the neighbouring mass distribution.

The main halo component is modelled considering a Navarro–Frenk–White \citep[NFW hereafter,][]{Navarro1997} profile. According to this model, the 3D radial density distribution of a halo can be described by:
\begin{equation} \label{eq:nfw}
\rho_{1 \rm h}(r) =  \dfrac{\rho_{\rm crit} \delta_{c}}{(r/r_{s})(1+r/r_{s})^{2}},
\end{equation}
where  $r_{s}$ is the scale radius and $\rho_{\rm crit}$ is the critical density of the Universe at the halo redshift. $\delta_{c}$ is the cha\-rac\-te\-ris\-tic overdensity
\begin{equation}
\delta_{c} = \frac{200}{3} \dfrac{c_{200}^{3}}{\ln(1+c_{200})-c_{200}/(1+c_{200})}.
\end{equation}
$c_{200} = r_{\Delta}/r_s$ is the halo concentration and $r_{200}$ is the radius that encloses a mean overdensity of $200 \times \rho_{\rm crit}$. The mass enclosed within $r_{200}$ is \mbox{$M_{200}=200\,\rho_{\rm crit} (4/3) \pi\,r_{200}^{3}$}. Thus, the whole mass distribution can be modelled by setting only two parameters: $M_{200}$ and $c_{200}$. 

The 2-halo term  affects mainly at larger radial scales introducing an excess in the halo mass distribution due to the neighbouring halos and depends on the mass of the main halo component, $M_{200}$. The 3D density distribution related with this term is obtained considering the halo-matter correlation function, $\xi_{\rm hm}$, as:
\begin{equation} \label{eq:rho2h}
    \rho_{2\rm h}(r) = \rho_{\rm m} \xi_{\rm hm} = \rho_{\rm crit} \Omega_m (1+z)^3 b(M_{200},\langle z \rangle) \xi_{\rm mm}
\end{equation}
where $\rho_m$ is the mean density of the Universe ($\rho_m = \rho_{\rm crit} \Omega_m (1+z)^3$) and the halo-matter correlation function is related with the matter-matter correlation function through the halo bias  \citep[$\xi_{\rm hm}  = b(M_{200},\langle z \rangle) \xi_{\rm mm}$,][]{Seljak2004}. We compute this by adopting \citet{Tinker2010} model calibration.

Both contrast density profiles, $\Delta \Sigma_{1 \rm h}$ and $\Delta \Sigma_{2\rm h}$, are obtained using \textsc{COLOSSUS}\footnote{\href{https://bitbucket.org/bdiemer/colossus/src/master/}{https://bitbucket.org/bdiemer/colossus/src/master/}} astrophysics toolkit \citep{Diemer2018}, which are computed by projecting the defined 3D density distributions, $\rho_{1 \rm h}$ (Eq. \ref{eq:nfw}) and $\rho_{2\rm h}$ (Eq. \ref{eq:rho2h}), respectively. Given that there is a well-known relation between the mass and the concentration, the lack of information in the inner regions of the density distribution could lead to biased results due to a poorly constrained concentration. Taking this into account, we adopt a similar approach as in other lensing studies \citep{Uitert2012, Kettula2015, Pereira2018, Gonzalez2021a} and fix the concentration using the relation with mass and redshift given by \citet{Diemer2019}. 

Computed profiles (Eq. \ref{eq:estimators}) are fitted from $300\,\text{kpc}/h$ to avoid the inner regions that could be affected by a stellar mass contribution of the central galaxies. Moreover, these regions are more affected by the background selection and the increased scatter due to the low sky area. The fitting procedure is implemented using \texttt{emcee} python package \citep{Foreman2013} that applies the Markov chain Monte Carlo (MCMC) method to optimise the log-likelihood $\ln{\mathcal{L}}(\Delta \Sigma(r) | M_{200})$. We first run 100 steps using 20 walkers to explore the parameter space considering a flat prior, $11.5 < \log(M_{200}/(h^{-1} M_\odot)) < 13.5$. Then, we run 1000 steps taking the final position of the walkers from the previous short run. Masses are obtained as the median values of the posterior distributions after discarding the first 125 steps of each chain, while errors enclose the $68\%$ of these distributions. 

\section{Results}
\label{sec:results}

\begin{figure*}
    \includegraphics[scale=0.6]{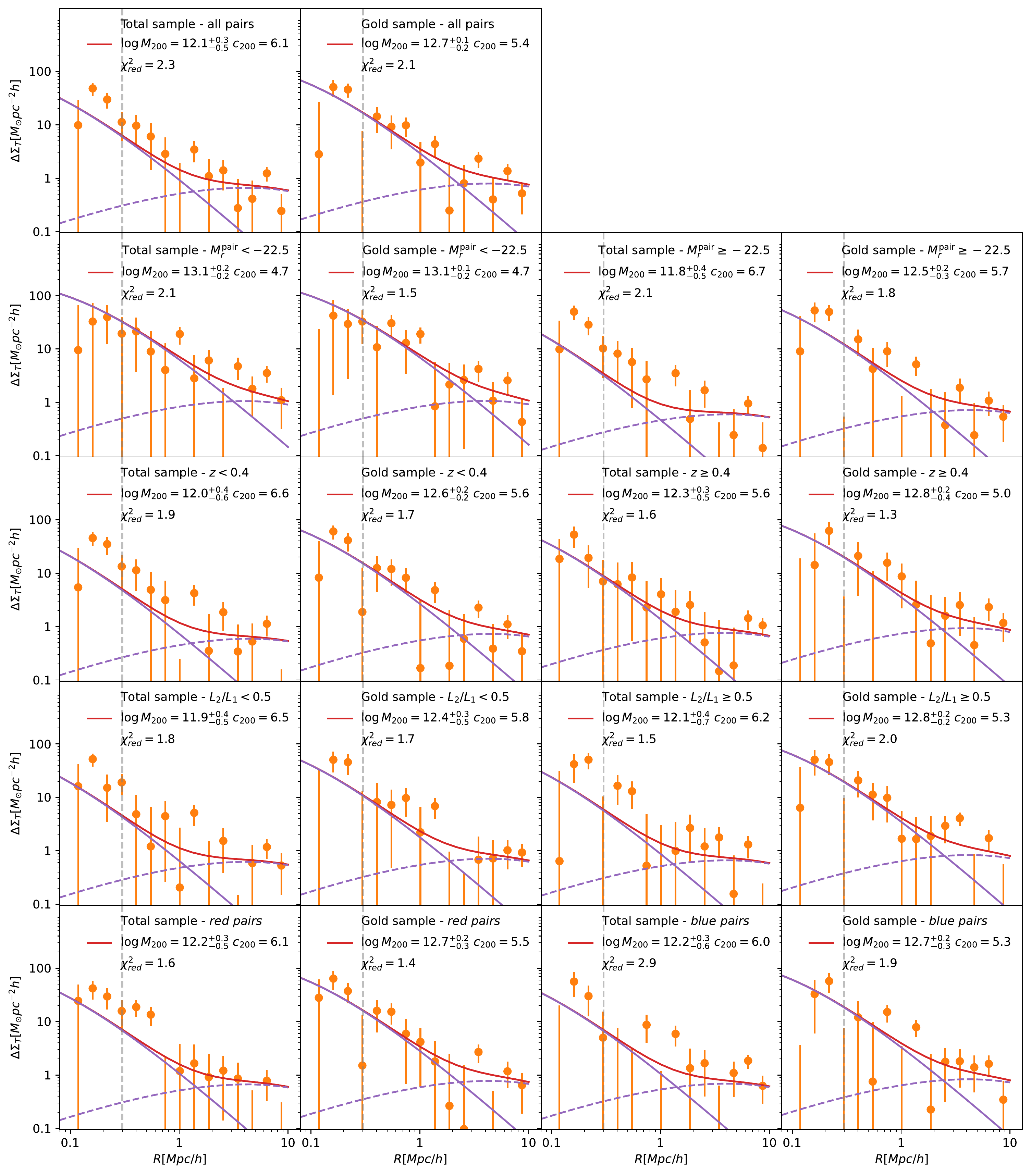}
    \caption{Computed radial contrast density profiles (orange dots) together with the total fitted model (Eq. \ref{eq:dsigma}, red solid line). The total model is computed considering the main halo component ($\Delta \Sigma_{1h}$, purple solid line) plus the neighbouring mass contribution component ($\Delta \Sigma_{2h}$, purple dashed line). In each panel, we specify the stacked sub-sample of pairs described in Table \ref{tab:sampdef}. In the legend we specified the fitted masses, $M_{200}$ in units of $M_\odot h^{-1}$, the correspondent concentrations and the reduced chi-squared values.}
    \label{fig:profile}
\end{figure*}

\subsection{Mass dependence on the galaxy pair properties}

We performed the lensing study for several galaxy pair sub-samples selected according to their main properties such as their total absolute magnitude, their luminosity ratio, colour and redshift (see Table \ref{tab:sampdef}). We show in Fig. \ref{fig:profile} the computed radial density contrast profiles for each sub-sample together with the fitted model, the obtained masses and concentrations and the reduced chi-square of the fit. For the complete sample of galaxy pairs identified using the \textit{Total} and the \textit{Gold sample}, we obtain mean masses of $10^{12.2} M_\odot h^{-1}$ and $10^{12.7} M_\odot h^{-1}$, respectively. Masses derived for the pairs identified using the \textit{Gold sample} of galaxies for all the sub-samples are in general higher, which is expected since it includes more luminous systems (Fig. \ref{fig:pair-props}).

\begin{figure}
    \includegraphics[scale=0.6]{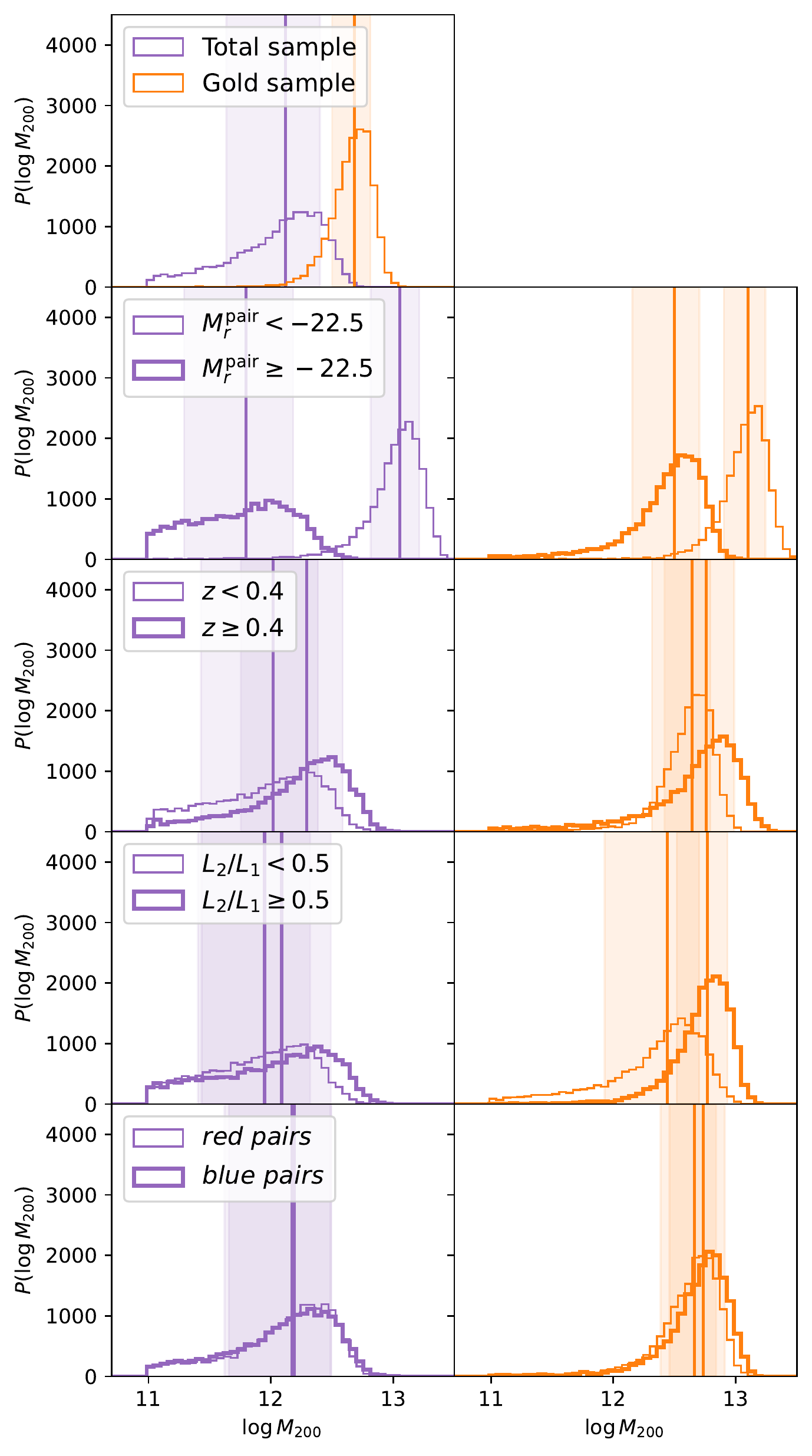}
    \caption{Posterior density distributions of the fitted $\log M_{200}$ after discarding the first 125 steps of each chain. We show the distributions for the pairs selected from the total and gold galaxy sub-samples in purple and orange, respectively. The selection cuts according to pair properties are shown in thicker and narrower lines as referred in the legends. Vertical lines indicate the median values and the shadow regions enclose 64$\%$ of the distributions.}
    \label{fig:pdd}
\end{figure}

In Fig. \ref{fig:pdd}, we show  the posterior density distributions obtained from the fitting procedure for all the analysed sub-samples. When considering lower luminosity pairs from the \textit{Total sample} ($M^\text{pair}_r \geq -22.5$) the derived mass is poorly constrained due to the low signal-to-noise ratio. We do not observe a mass dependence with redshift for the pairs selected using the \textit{Total sample}. On the other hand, pairs located at higher redshifts identified from the \textit{Gold sample} of galaxies tend to show slightly higher mean masses. This result is related to a larger number of brighter galaxies at higher redshifts. When selecting the pairs according to the luminosity ratio, we obtain that pairs with components that have similar luminosity ($L_2/L_1 \geq 0.5$) show higher masses than those with a dominant component, for which the mass is poorly constrained. Finally, we do not obtain a noticeable mass dependence on colour. 

\subsection{Galaxy pair mass-luminosity relation}
\label{subsec:mlrelation}

\begin{figure*}
    \includegraphics[scale=0.6]{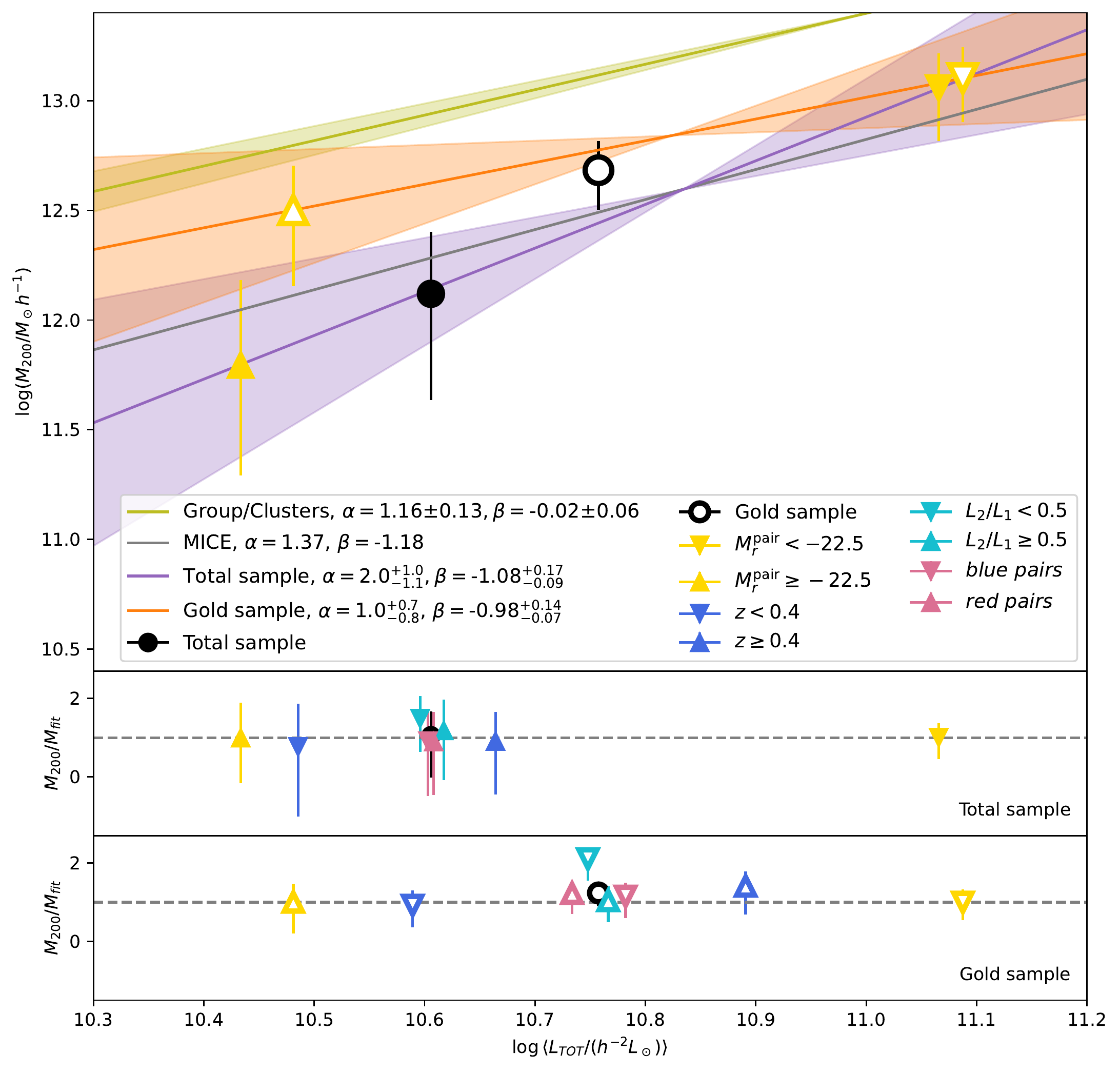}
\caption{\textit{Upper panel:} Lensing mass estimates compared with the mean total luminosity of the stacked pairs, $L_\text{TOT} = L_1 + L_2$, $\langle M_r \rangle$. Purple and orange solid lines correspond to the linear fits of the pairs selected according to the total absolute magnitude (yellow markers), from the \textit{Total} and \textit{Gold} galaxy sub-sets, respectively. Shadow regions correspond to the errors in the fitted data. Fitted slope, $\alpha$, and intercept, $\beta$, are specified in the legend. The green solid line and shadow region correspond to the mass-to-light relation derived from spectroscopically selected galaxy groups and clusters \citep{Viola2015}. \textit{Middle panel}: Ratio between derived masses for all the pair sub-sets selected from the \textit{Total sample} and the fitted relation. Colour code and symbols represent the sub-samples described in Table \ref{tab:sampdef}. Horizontal dashed grey line is set to unity. \textit{Lower panel:} Same as for the middle panel but for the galaxy pairs selected from the \textit{Gold sample}.}
    \label{fig:masslight}
\end{figure*}

To better understand the relation between the considered pair properties and the halo masses, we compare the obtained lensing masses, $M_{200}$, with the mean total luminosity in the $r-$band of the stacked pairs in each sub-sample, $\langle L_1 + L_2 \rangle$. We fit a power-law relation between these parameters defined as:
\begin{equation}
\frac{M_{200}}{10^{14} M_\odot h^{-1}} = \beta \left( \frac{\langle L_1 + L_2 \rangle}{10^{10.5}h^{-2}L_\odot} \right)^\alpha,
\end{equation}
where $L_\odot$ is the solar luminosity in the $r-$band. The fitting procedure is performed on the log-basis using least square minimisation and taking into account the mean of the errors in the mass estimates. We take into account only the two sub-sets selected according to the total magnitude, those with $M^\text{pair}_r \geq -22.5$ and those with $M^\text{pair}_r < -22.5$, obtained from \textit{Gold}  and \textit{Total sample} of galaxies separately. Fitted relations and obtained parameters are shown in Fig. \ref{fig:masslight}. In general, the results for all the sub-samples are in agreement with the fitted relations. Pair sub-samples selected from the \textit{Total sample} predict a steeper mass-to-light relation than the relation obtained using the pairs identified from the \textit{Gold sample}, and are in good agreement within $1\sigma$ errors with the expected mass-luminosity ratio derived from MICE simulated data in \citet{Rodriguez2020}.

When galaxy pairs are selected according to their colour or luminosity ratio, the resulting masses are in excellent agreement with the ones derived for the whole sample of pairs. A slight difference is obtained for the pairs with low luminosity ratios ($L_2/L_1 < 0.5$) identified using the \textit{Gold sample}. Although this subset of pairs includes a lower mean luminosity than those selected according to $L_2/L_1 \geq 0.5$, we obtain a higher mass estimate. This might be indicating that pairs with a dominant galaxy component are on average hosted in more massive halos than those in which the components share a similar luminosity. 

We also include for comparison the mass-to-light relation derived by \citet{Viola2015}, based on a lensing analysis of a sample of spectroscopically selected galaxy groups and clusters. Taking into account that this relation is obtained for systems more massive than the analysed in this work ($\sim 10^{13} - 10^{14.5} h^{-1} M_\odot$), the study presented here allows us to test this relation at the lower mass regions. In order to include this relation, we scale their estimated $\beta$ value given that we choose a different luminosity pivot to fit the parameters. We found a good agreement with the slope obtained by \citet{Viola2015}, but our relation systematically predicts lower masses. 

The study of the relation between the mass and light through the mass-to-light ratio of individual galaxies and galaxy systems, contributes to the general understanding of the connection between the dark and the luminous components, i.e. between the galaxies and their host dark matter halos. According to our current understanding of galaxy formation, it is expected a change in the slope of the mass–luminosity relation towards a mass range of $10^{12} - 10^{13} M_\odot h^{-1}$, given that star formation is more efficient in
halos of $\sim 10^{12} M_\odot h^{-1}$ \citep{Bardeen1986,davis1985,Girardi2002,Behroozi2013}. In particular, pairs identified in this work are constrained within the luminosity and mass ranges in which the change in the slope of the mass-to-light relation is predicted. We found a good agreement between our mass-to-light ratios for the galaxy pairs and those found for late-type galaxies presented in \citet{Mandelbaum2006}, in which they obtain an almost plane relation with a mean $M_{200}/L_r \sim 80 \, M_\odot/(h\,L_\odot).$\footnote{For this comparison we transform the masses derived by \citet{Mandelbaum2006} using the median density, $200\bar{\rho}$, to our mass definition in terms of the critical density.}

A roughly flat relation ($\alpha\sim1$) is obtained for the pairs identified from the \textit{Gold sample} that corresponds to a total mass-luminosity ratio of $M_{200}/(L_1+L_2) \sim 100 M_\odot/(h\,L_\odot)$. However, a steeper relation ($\alpha\sim2$) is derived for the pairs identified from the \textit{Total sample}, where $M_{200}/(L_1+L_2) \sim 23\, M_\odot/(h\,L_\odot)$ and $M_{200}/(L_1+L_2) \sim 98\, M_\odot/(h\,L_\odot)$ is found for the pair sub-sets with $M^\text{pair}_r \geq -22.5$ and $M^\text{pair}_r < -22.5$, respectively. Pairs identified from the \textit{Gold sample} are expected to have a larger chance of including systems with a less bright non-detected component besides the identified members, due to the cut applied according to the redshift photometric quality in which dimmer objects can be discarded. This, in turn, can bias the obtained relation given that the total luminosities might be underestimated, thus the obtained flatter relation for these subsets can be related with brighter systems with a slope more in agreement with the derived for group halo scales.


\section{Summary and conclusions}
\label{sec:conclusion}

In this work, we presented a galaxy pair catalogue based on PAUS data. Galaxy pairs are identified according to the projected distance and the difference in the radial velocity of the galaxies, computed using the high-quality photometric redshifts provided by the survey. By applying the algorithm presented in \citet{Rodriguez2020}, we identify the systems using two sub-sets of galaxies within $0.2 < z <0.6$ and with an apparent magnitude $i < 23$. The two subsets are the full sample of galaxies provided by PAUS after discarding stars and spurious objects (\textit{Total sample}) and a more restrictive sample that includes only galaxies with high-quality redshift estimates (\textit{Gold sample}). Galaxy pairs identified from the \textit{Gold sample} are in general more luminous and located at lower redshifts than those identified in the \textit{Total sample}. 

Identified pairs were divided into sub-samples selected according to their properties, such as colour, total absolute magnitude and redshift, and we measured the mean masses of the pairs included in each sub-sample by applying weak-lensing stacking techniques.  We do not observe a redshift dependence on the estimated masses. This indicates that the identification is stable with redshift, i.e. the identified systems share mean similar properties regardless of the redshift at least within the considered range. We found a slight difference in the lensing masses when the pairs are selected according to their component luminosity ratio, specially for the pairs obtained from the \textit{Gold sample}. Our results indicate that pairs with a dominant luminous component ($L_2/L_1 < 0.5$) are on average more massive systems than those in which the components
share a similar luminosity. Also, we do not observe a dependence between the mass and the pair colour. 

We also provide a parametric relation between the pair luminosity and their expected mass, derived by considering the pairs sub-samples selected by their absolute magnitude. We obtain a steeper relation for the \textit{Total sample}, in agreement with the one presented in \citet{Rodriguez2020} using simulated data, than for the \textit{Gold sample}. The lower slope obtained using the pairs from the \textit{Gold sample} can be related with an underestimated total luminosity, since the applied cut according to the photo$-z$ quality tends to discard dimmer objects that can be part of the identified systems. The resultant fitted relations were compared with previous mass-to-light ratio studies for individual galaxies, galaxy groups and clusters. Our results for the galaxy pairs are mainly in agreement with the mass-to-light ratio derived for late-type galaxies, while our masses are systematically lower than the one predicted by the relations obtained for groups and clusters. 
 
This study provides a sample of these particular galaxy systems that are of special interest in galaxy formation and evolution and constitute one of the initial bricks in the halo assembly. We provide the mean masses that characterise the identified systems and analyse their physical properties. This study allowed also to constrain the mass-luminosity relation of galaxy systems at the lower mass regions. We particularly highlight that the present study is based on photometric redshift estimates, validating the potential usage of these measurements in this new era of photometric wide-field surveys and allowing a higher completeness of the identified systems. Moreover, since spectroscopic catalogues at this limiting magnitudes ($i < 23$) are obtained by targeting galaxies with particular photometric properties (e.g. VIPERS, COSMOS), the released catalogue, in particular from the \textit{Total sample}, ensures a higher completeness in the identified members allowing a better characterisation of the pairs.

\section*{Acknowledgements}
We kindly thank Jon Loveday for his useful comments that help to improve this manuscript. 
The PAU Survey is partially supported by MINECO under grants CSD2007-00060, AYA2015-71825, ESP2017-89838, PGC2018-094773, PGC2018-102021, PID2019-111317GB, SEV-2016-0588, SEV-2016-0597, MDM-2015-0509 and Juan de la Cierva fellowship and LACEGAL and EWC Marie Sklodowska-Curie grant No 734374 and no.776247 with ERDF funds from the EU Horizon 2020 Programme, some of which include ERDF funds from the European Union. IEEC and IFAE are partially funded by the CERCA and Beatriu de Pinos program of the Generalitat de Catalunya. Funding for PAUS has also been provided by Durham University (via the ERC StG DEGAS-259586), ETH Zurich, Leiden University (via ERC StG ADULT-279396 and Netherlands Organisation for Scientific Research (NWO) Vici grant 639.043.512), University College London and from the European Union's Horizon 2020 research and innovation programme under the grant agreement No 776247 EWC. The PAU data center is hosted by the Port d'Informaci\'o Cient\'ifica (PIC), maintained through a collaboration of CIEMAT and IFAE, with additional support from Universitat Aut\`onoma de Barcelona and ERDF. We acknowledge the PIC services department team for their support and fruitful discussions.
This project has received funding from the European Union’s Horizon 2020 Research and Innovation Programme under the Marie Sklodowska-Curie grant agreement No 734374. This work was also partially supported by the Consejo Nacional de Investigaciones Científicas y Técnicas (CONICET, Argentina)
and the Secretaría de Ciencia y Tecnología de la Universidad Nacional de Córdoba (SeCyT-UNC, Argentina). This work has made use of CosmoHub. CosmoHub has been developed by the Port d’Informació Científica (PIC), maintained through a collaboration of the
Institut de Física d’Altes Energies (IFAE) and the Centro de Investigaciones Energéticas, Medioambientales y Tecnológicas (CIEMAT), and was partially funded by the “Plan Estatal de Investigación Científica y Técnica y de Innovación” program of the Spanish government.
This work has been also partially supported by the Polish National Agency for Academic Exchange (Bekker grant BPN/BEK/2021/1/00298/DEC/1), the European Union's Horizon 2020 Research and Innovation programme under the Maria Sklodowska-Curie grant agreement (No. 754510).
H. Hildebrandt is supported by a Heisenberg grant of the Deutsche Forschungsgemeinschaft (Hi 1495/5-1) as well as an ERC Consolidator Grant (No. 770935).
A. Wittje is supported by the DFG (SFB 1491).

\section*{Data Availability}

The PAUS data underlying this article will be shared on reasonable request to the corresponding authors. The released galaxy pair catalogue can be acquired from \href{https://cosmohub.pic.es/catalogs/309}{https://cosmohub.pic.es/catalogs/309}.



\bibliographystyle{mnras}
\bibliography{references} 




\appendix

\section{The galaxy pair catalogue}
\label{app:columns}

We specify the columns of the released catalogue (\href{https://cosmohub.pic.es/catalogs/309}{https://cosmohub.pic.es/catalogs/309}) in Table \ref{tab:columns}.

\begin{table*}
\caption{Description of the galaxy pair catalogue columns.}
\begin{center}
\begin{tabular}[c]{l l l l}
 \hline
Column name & Format & Description & Units \\
\hline\hline
PAUS\_sample & tinyint & galaxy sample from which the pairs were identified [0 = Total, 1 = Gold] & \\ 
pair\_id & int & pair id number & \\
ra\_1 & float & right ascenction of the brightest pair component & (deg) \\
dec\_1 & float & declination of the brightest pair component & (deg) \\
zphot\_1 & float & photometric redshift of the brightest component &  \\
mag\_i\_1 & float & $i-$band apparent magnitude for the brightest component &  \\
mag\_r\_1 & float & $r-$band apparent magnitude for the brightest component &  \\
ra\_2 & float & right ascenction of the faintest pair component & (deg) \\
dec\_2 & float & declination of the faintest pair component & (deg) \\
zphot\_2 & float & photometric redshift of the faintest component &
\\
mag\_i\_2 & float & $i-$band apparent magnitude for the brightest component &  \\
mag\_r\_2 & float & $r-$band apparent magnitude for the brightest component &  \\

\hline
\end{tabular}
\end{center}
\label{tab:columns}
\end{table*}


\bsp	
\label{lastpage}
\end{document}